\documentclass{aa}
\usepackage[varg]{txfonts}

\usepackage{array, multirow, graphicx,natbib,txfonts,color,tabularx,amsmath}
\usepackage{lscape,float,placeins}
\usepackage[dvipsnames]{xcolor} \usepackage[colorlinks=true,allcolors=blue,urlcolor=CadetBlue,breaklinks=true]{hyperref}
\usepackage{afterpage}
\usepackage{threeparttable}
\bibpunct{(}{)}{;}{a}{}{,}

\usepackage{xspace}

\newcommand{\ang}{\ensuremath{\text{\AA}}\@\xspace}

\newcommand{\XY}[2]{\ensuremath{\left[\mbox{#1/#2}\right]}}
\newcommand{\FeH}{\XY{Fe}{H}}
\newcommand{\eFe}[1]{\XY{#1}{Fe}}
\newcommand{\eH}[1]{\XY{#1}{H}}
\newcommand{\alphaFe}{\XY{$\alpha$}{Fe}}

\newcommand{\kms}{\text{km\,s}^{-1}}
\newcommand{\Teff}{{T_{\mbox{{\scriptsize eff}}}}}
\newcommand{\logg}{{\log g}}

\newcommand{\vmic}{{v_{\mbox{{\scriptsize mic}}}}}

\newcommand{\Abund}[1]{A(\text{#1})}

\newcommand{\conf}[2]{$\rm #1 \: #2$\@\xspace}

\newcommand{\stagger}{\textsc{stagger}\@\xspace}

\newcommand{\marcs}{{MARCS}\@\xspace}
\newcommand{\falc}{{FALC}\@\xspace}
\newcommand{\multi}{\textsc{multi}\@\xspace}
\newcommand{\multitd}{\textsc{multi3d}\@\xspace}
\newcommand{\scate}{\textsc{scate}\@\xspace}
\newcommand{\avgtd}{$\langle3\text{D}\rangle$\@\xspace}

\newcommand{\BSMS}{HD~22879\@\xspace}
\newcommand{\BSTOP}{HD~84937\@\xspace}
\newcommand{\BSSGB}{HD~140283\@\xspace}
\newcommand{\BSRGB}{HD~122563\@\xspace}

\newcommand*{\eg}{e.g.\@\xspace}
\newcommand*{\ie}{i.e.\@\xspace}

\usepackage{upgreek} \newcommand\micron{\ensuremath{\rm \upmu m}\@\xspace} 

\begin{document}

\title{Non-LTE aluminium abundances in late-type stars}

\author{T.~Nordlander\inst{1,2} \and K.~Lind\inst{1,3}}

\offprints{thomasn@mso.anu.edu.au}

\institute{Division of Astronomy and Space Physics, Department of Physics and Astronomy, Uppsala University, PO Box 516, 751 20 Uppsala, Sweden
	\and Research School of Astronomy and Astrophysics, Australian National University, ACT 2611, Australia. \\ \email{thomasn@mso.anu.edu.au}
	\and Max-Planck-Institut f\"ur Astronomie, K\"onigstuhl 17, 69117, Heidelberg, Germany 
}

\date{Received 12 January 2017 / Accepted 28 July 2017}

\authorrunning{T. Nordlander \& K. Lind}
\titlerunning{NLTE aluminium abundance analyses}

\abstract
{}
{
Aluminium plays a key role in studies of the chemical enrichment of the Galaxy and of globular clusters. However, strong deviations from LTE (non-LTE) are known to significantly affect the inferred abundances in giant and metal-poor stars.}
{We present NLTE modeling of aluminium using recent and accurate atomic data, in particular utilizing new transition rates for collisions with hydrogen atoms, without the need for any astrophysically calibrated parameters. For the first time, we perform 3D NLTE modeling of aluminium lines in the solar spectrum. We also compute and make available extensive grids of abundance corrections for lines in the optical and near-infrared using one-dimensional model atmospheres, and apply grids of precomputed departure coefficients to direct line synthesis for a set of benchmark stars with accurately known stellar parameters.}
{Our 3D NLTE modeling of the solar spectrum reproduces observed center-to-limb variations in the solar spectrum of the 7835\,\AA\ line as well as the mid-infrared photospheric emission line at 12.33\,$\mu$m. 
We infer a 3D NLTE solar photospheric abundance of $\Abund{Al} = 6.43 \pm 0.03$, in exact agreement with the meteoritic abundance. 
We find that abundance corrections vary rapidly with stellar parameters; for the 3961\,\AA\ resonance line, corrections are positive and may be as large as +1 dex, while corrections for subordinate lines generally have positive sign for warm stars but negative for cool stars. 
Our modeling reproduces the observed line profiles of benchmark K-giants, and we find abundance corrections as large as $-0.3$\,dex for Arcturus. 
Our analyses of four metal-poor benchmark stars yield consistent abundances between the 3961\,\AA\ resonance line and lines in the UV, optical and near-infrared regions. Finally, we discuss implications for the galactic chemical evolution of aluminium. }
{}


\maketitle


\section{Introduction}\label{sect:intro}
The formation and evolution of the Galaxy can be traced by analysing the ages, kinematics and chemistry of late-type stars. 
Aluminium plays a key role in studies of the chemical enrichment of the Galactic halo and disk, as well as in stellar clusters.
As the synthesis of the only stable nucleus $^{27}$Al requires an excess of neutrons, the yield is sensitive to the initial composition and mass of the polluting star. 
In massive stars that explode as core-collapse supernovae (SNe II) or pair instability supernovae (PISN), this neutron excess is provided by the neutron-rich isotope $^{22}$Ne, which in turn is a result of He-burning of nitrogen created in the CNO cycle \citep{kobayashi_galactic_2006}. 
In zero-metallicity population III stars, the seed metals instead come from primary production of carbon, resulting in strongly mass dependent Al yields for SNe II \citep[see Fig.~10 of][]{heger_nucleosynthesis_2010}, but largely constant in PISN \citep{heger_nucleosynthetic_2002}.
In population II stars, the neutron excess is metallicity dependent \citep[see \eg][]{kobayashi_galactic_2006,kobayashi_chemodynamical_2011}, contrary to the production of \eg $\alpha$-elements.
Additionally, the SNe II yield of Al is sensitive to details of the explosion or mixing mechanism \citep{iwamoto_first_2005}.

The observational manifestation of the SNe II yields is debated. \citet{gehren_na_2006} demonstrated a clear bimodality in $\XY{Al}{Mg}$ with a separation of at least 0.2\,dex between disk and halo stars, with no dependence on $\eFe{Mg}$, and highlight the strong sensitivity of the abundance ratio to departures from local thermodynamic equilibrium (LTE). 
The $\XY{Al}{Mg}$ ratio in the Galactic disk itself is found to be essentially constant, because SNe Ia produce very little of either element \citep[\eg][]{nomoto_nucleosynthesis_1997}. However, the large Fe-yield of SNe Ia makes the $\eFe{Al}$ ratio an excellent diagnostic alongside $\alphaFe$ to separate the thin and thick disk \citep[\eg][]{bensby_exploring_2014}.

Finally, stars in globular clusters exhibit (anti-)correlated abundance variations in aluminium and other light elements as large as 1\,dex \citep[\eg][]{carretta_na-o_2009}.
The stars with enhanced abundances of odd-Z elements such as N, Na and Al are thought to have formed in a second (or third) burst of star formation, after the interstellar medium in the cluster was polluted by massive first generation stars \citep[see \eg the reviews by][]{kraft_abundance_1994,gratton_multiple_2012}.
The enrichment process is proton capture in either the CNO-cycle, or the NeNa- or MgAl-chains, which operate at different temperatures \citep[see \eg the review by][]{charbonnel_multiple_2016}.
The sites of these processes have been suggested to be, \eg, asymptotic giant branch stars (AGB) of low \citep{ventura_predictions_2001} or intermediate mass \citep{ventura_hot_2011}, or in fast rotating massive stars \citep{decressin_fast_2007}.
The extent of the abundance variations in different elements and the relative numbers of first and second generation stars may be used to determine the nature of the polluting stars, the timescale of the star formation episodes, and the initial mass of the stellar cluster \citep{carretta_properties_2010}.

In all these scenarios, the key to understanding the underlying physical processes is an accurate and reliable determination of chemical abundances. 
Aluminium has long been known to be problematic in this aspect, as departures from LTE (NLTE) have been found empirically in abundance comparisons between dwarfs and giants in the Galactic disk \citep{begley_evidence_1987}, as well as theoretically in several studies using now outdated atomic data \citep[\eg][]{gehren_spectroscopic_1991,baumuller_line_1996,andrievsky_nlte_2008}. 
In particular the recent availability of accurate calculations describing inelastic collisions with hydrogen atoms \citep{belyaev_inelastic_2013} removes one of the main shortcomings in these analyses, which relied on the classical formula of \citet{drawin_zur_1968} with an empirical scaling factor typically determined from analyses of the solar spectrum.

We present in Sect.~\ref{sect:methods} a new atomic model of aluminium using the latest data for radiative and collisional transitions, which we apply to calculations using both classical \marcs hydrostatic model atmospheres \citep{gustafsson_grid_2008} and \avgtd hydrodynamical \stagger models \citep{magic_stagger-grid:_2013}. 
We compute extensive grids of NLTE corrections, and apply them in Sect.~\ref{sect:results} to the solar flux and intensity spectra as well as to the spectra of late-type standard stars at high and low metallicity in order to assess their accuracy. 
Finally, the NLTE grids are presented (Sect.~\ref{sect:hrdiag}) and compared to previous literature (Sect.~\ref{sect:previous}).



\section{Methods}\label{sect:methods}

\subsection{NLTE modeling}
We solve the statistical equilibrium in 1D model atmospheres using the \multi code \citep[version 2.3;][]{carlsson_computer_1986,carlsson_multi_1992}. Calculations of bound-free transitions use background line opacities that are consistent with those used in the \marcs grid of model atmospheres \citep{gustafsson_grid_2008}. 
The resonance lines at 3944 and 3961\,\ang lie between the strong \ion{Ca}{ii} H and K lines as well as Balmer-$\varepsilon$, and we include these explicitly in our calculations.
We use the grid of temporally and spatially averaged (\avgtd) hydrodynamical \stagger models \citep{magic2013_grid}, and adopt $\vmic = 1\,\kms$ for dwarfs and subgiants ($\logg \ge 3.5$) and $\vmic = 2\,\kms$ for giants ($\logg \le 3$) in our NLTE calculations.
We also use 1D hydrostatic model atmospheres from the \marcs grid \citep{gustafsson_grid_2008} which were computed in plane-parallel geometry with $\vmic = 1\,\kms$ for dwarfs ($\logg \ge 4$), and in spherically symmetric geometry with $\vmic = 2\,\kms$ for giants and subgiants ($\logg \le 3.5$), and we use these values consistently in our NLTE calculations.

For the Sun, we perform full 3D NLTE calculations using \multitd \citep{leenaarts_multi3d_2009}.
The \multitd code is used as described by \citet{amarsi_non-lte_2016-1}, with background opacitites computed for this work. 
We solve the statistical equilibrium using 26 short characteristic rays \citep[see][]{amarsi_solar_2017} for a range of abundances using five snapshots taken from an updated version \citep[see][]{lind_non-lte_2017} of the solar radiation hydrodynamical simulation used by \citet{scott_elemental_2015-1} but resampled from the original resolution of $240^2 \times 230$ to $60^2 \times 101$ (horizontal $\times$ vertical).
We also compute LTE line profiles in a range of abundances for a larger set of 15 snapshots using \scate \citep{hayek_3d_2011}, and apply the NLTE/LTE profile ratios computed with \multitd at each viewing angle $\mu$ and each abundance to these LTE line profiles.

Finally, for the Sun, we also test the influence of a 1D chromospheric temperature structure using the FALC model \citep[model ``C'' of][]{fontenla_energy_1993} using the same setup as for 1D \marcs and \avgtd \stagger models.

\begin{figure}
	\includegraphics[width=.48\textwidth,clip,trim=0 2.3em 1em 3em]{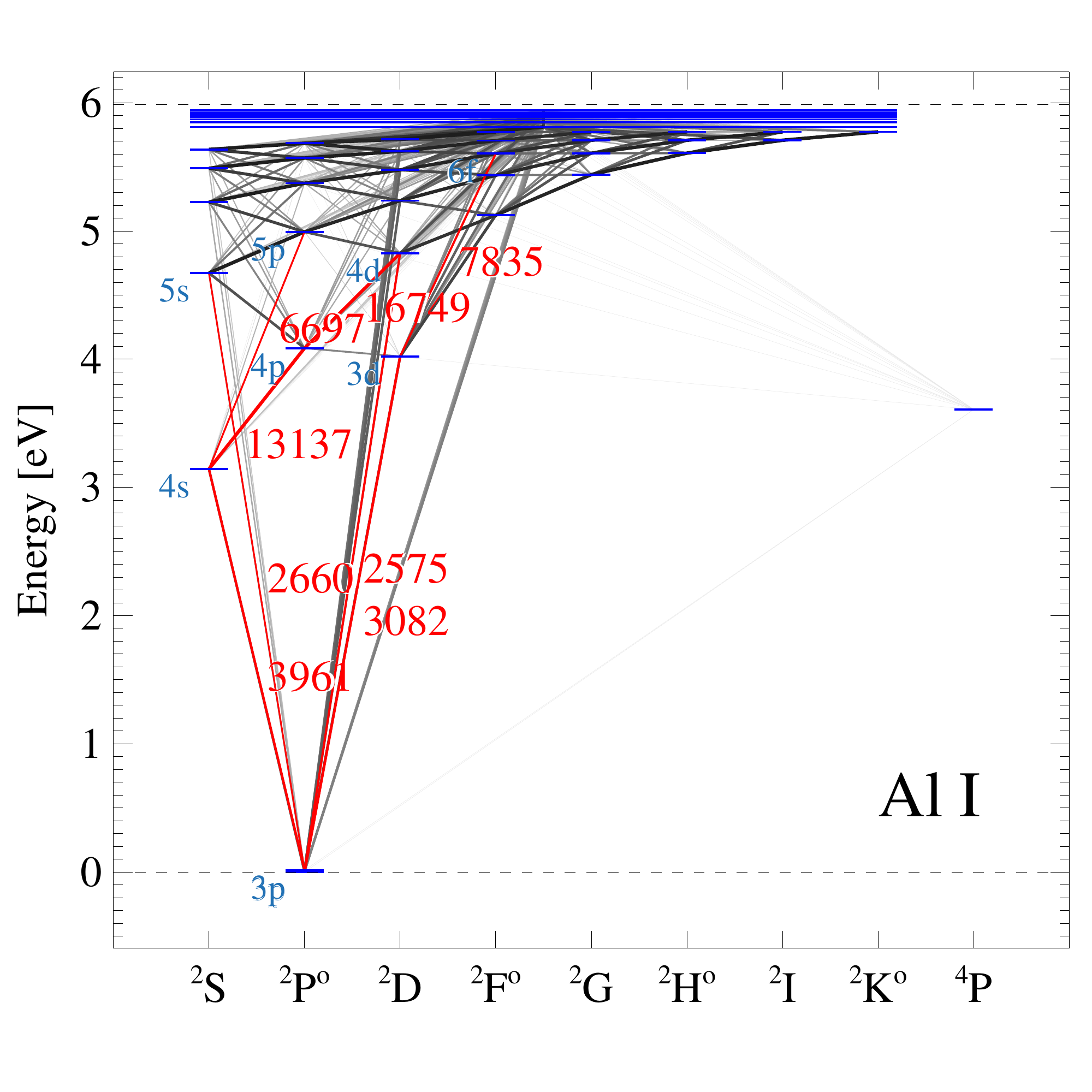}
	\caption{Term diagram with bound-bound transitions considered in the solution of the statistical equilibrium. Levels of \ion{Al}{ii} are not shown. Darker colors indicate a larger $f$-value. Important lines (multiplets) used in the abundance analysis are shown in red and labeled. }
	\label{fig:termdiag}
\end{figure}
\subsubsection{Energy levels}
Aluminium is a rather simple atom in the boron group, with ground configuration \conf{2p^6\, 3s^2\, 3p}{^2P^o}. 
All bound states have a single excited electron outside the $\rm 3s^2$ shell, giving doublet configurations of the type $3\text s^2\,nl\:^2L$, except the double-excited state \conf{3 \text s 3\text p^2}{^4\text P}.

We adopt energy levels from NIST \citep{martin_energy_1979,kelleher_atomic_2008}, but note an inconsistency in \ion{Al}i. As reviewed by \citet{martin_energy_1979}, \conf{3s3p^2}{^2D} does not exist as a localized state, but instead perturbs the $3{\rm s}^2 n {\rm d}\:^2{\rm D}$ series so that in particular lower members deviate from the expected $n {\rm d}$ orbitals. To emphasize this, \citet{kelleher_atomic_2008} and many laboratory publications denote the $n {\rm d}$ series by 3d, $n {\rm d}$, 4d, 5d, \ldots, while \eg the VALD, TOPbase and Kurucz databases name series members by the standard 3d, 4d, 5d, 6d, \ldots\ scheme. We use this standard series naming scheme for consistency with the astronomical databases.

For high $n$-values of \ion{Al}i, we supplement the NIST data with theoretical predictions from \citet[updated in 2012\footnote{Available online at \url{http://kurucz.harvard.edu/atoms/1300/}}]{kurucz_kurucz_1995}, and use hydrogenic energies otherwise.

We resolve the ground states as two components, but otherwise average fine-structure levels into a single state with a $g$-weighted energy. 
Our description of \ion{Al}i is truncated at $n=20$, and we combine states with different $l$ into superlevels for $n=9$--15. All levels with $n=16$--20 are combined into a single superlevel. This results in 42 levels of \ion{Al}i. For \ion{Al}{ii}, we retain only the ground \conf{3s^2}{^1S} and first excited \conf{3s3p}{^3P^o} state.

The completeness of the atom was tested on a set of metal-rich and metal-poor atmospheric models representative of the Sun, F-stars and K-giants, where we required that the populations of levels involved in diagnostic lines agree to within 1\,\% in the line forming regions between the reduced atom, and a comprehensive atom. The comprehensive atom was represented by 72 states of \ion{Al}i where states with $n=15$--20 are combined into superlevels, 36 states of \ion{Al}{ii} complete up to $n=6$, and the \ion{Al}{iii} ground state.

\subsubsection{Radiative transitions}

We use oscillator strengths from NIST \citep{kelleher_atomic_2008}, which are mainly those of \citet{tachiev_almchf_2002}
for low-energy states and TOPbase \citep{butler_atomic_1993,mendoza_atomic_1995} otherwise. 
We supplement these with data from Kurucz, and use hydrogenic transition probabilities otherwise.
For the $\rm 6h$--$\rm 7i$ transition at 12.33\,\micron, we adopt the wavelength from \citet{chang_identification_1983}, which is 7\,\% shorter than that computed by Kurucz. As $f \propto \lambda^{-1}$, we correct the transition probability accordingly.

For broadening due to collisions with hydrogen we interpolate ABO data published in a series of papers \citep{anstee_width_1995,barklem_broadening_1998,barklem_broadening_1997} using the \textsc{abo-cross} code \citep{barklem_abo-cross:_2015}. 
We also utilize new broadening data computed for this work (provided by P.\,S. Barklem) using ABO theory when only low-excitation states are involved, or otherwise the impact approximation \citep[see Sect.~3.2.1 of][]{osorio_mg_2015-1}. 
Thus, every line of aluminium analysed in this work that is significantly pressure broadened utilizes realistic broadening parameters.
For other lines, we use the \citet{unsold_physik_1955} approximation with an enhancement factor of 2.5. 
We compute line profiles using the line data given in Table~\ref{tbl:linedata}.

We neglect hyperfine splitting (HFS) in the solution of the statistical equilibrium, but take it into account in our spectrum calculations. We follow the review by \citet{chang_energy_1990}, but update the HFS constants using more recent literature when available, and summarise our selected values in Table~\ref{tbl:hfs}. 
Constants are extrapolated as $A \propto n_\text{eff}^{-\alpha}$ where $n_\text{eff}$ is the effective quantum number, with varying values of the exponentials $\alpha$ suggested by \citet{chang_energy_1990}. While the lower members of the $^2D_{5/2}$ series are significantly broadened, it is not possible to predict the HFS constants of higher states due to the strong perturbation by \conf{3s3p^2}{^2D} mentioned in the previous section.

Photoionization cross-sections come from TOPbase \citep{cunto_topbase_1993-1}. For \ion{Al}i, these are complete up to $nl=9\text{h}$, and also include the $nl = 10$s--d and 11s--p states, as well as the $\rm 3s 3p^2$ state.
We use hydrogenic approximations for other states. 
We neglect b-f transitions for \ion{Al}{ii}, as the ionization thresholds of the most populated states lie blueward of the Lyman limit.

The TOPbase cross-sections are tabulated on energy grids which are sufficiently dense to resolve the narrow and sometimes rather chaotic resonance features. Due to uncertainties in the calculated electronic structure, the positions of these resonances are however rather uncertain, and errors may cause them to spuriously coincide with important background opacity features. \citet{bautista_resonance-averaged_1998} argued that this problem can be circumvented by smoothing the cross-sections to represent the estimated uncertainties in energy levels. \citet{allende_prieto_non-lte_2003} have published such smoothed cross-sections as part of their NLTE model atom database, and we use their data after interpolating them onto equidistant frequency grids. We find that a sparse sampling of 600\,$\kms$ ($\lambda / \delta \lambda = R = 500$) is sufficient for the first few excited states whose ionization thresholds lie in the optical, while lower resolution is sufficient for more highly excited states with thresholds in the infrared.

\subsubsection{Collisional transitions}
Previous work on aluminium was hampered mainly by uncertainties in hydrogen collisional rates. Semi-empirical recipes based on the formalism of \citet{drawin_zur_1968} as implemented by \citet{steenbock_statistical_1984} were typically used, although these rates are known to generally be too large, with estimates ranging from factors of a few to factors of thousands.
\citet{baumuller_line_1996,baumuller_aluminium_1997} used a scaling factor $S_\text H = 0.002$ for hydrogen collision rates based on analyses of the solar emission features near 12\,\micron (involving high-excitation, Rydberg, states), while a higher factor $S_\text H = 0.4$ was required to bring abundances determined from different optical lines (involving low-excitation states) into agreement. Other works have adopted similar values for low-excitation lines, \eg $S_\text H = 0.1$ \citep{andrievsky_nlte_2008} and $S_\text H = 1/3$ \citep{steenbock_statistical_1992}. 

We adopt realistic collisional rates for the first six low-excitation states as computed by \citet{belyaev_inelastic_2013} using a Landau-Zener-like approach \citep{belyaev_model_2013}. These collisional excitation rates are significantly lower than those estimated with the Drawin formula by roughly two orders of magnitude. In particular, the Drawin formula predicts large excitation rates from the ground state, while the quantum mechanical approach finds them to be completely negligible. 
Additionally, \citet{belyaev_inelastic_2013} compute charge transfer rates for which no classical recipes exist. Transition rates for the low-lying excited states are very large, introducing an efficient thermalizing mechanism which was not predicted by the Drawin formula.
These transition rates have previously been used in the work of \citet{mashonkina_influence_2016}. They found that in the solar photosphere, excitation rates due to collisions with electrons were much higher than those with hydrogen atoms for transition energies greater than 3\,eV, but comparable for transition energies smaller than 2\,eV.

For Rydberg states, \ie, all states except the first six, we compute hydrogen impact transition rates using the free electron model \citep{kaulakys1985,kaulakys1991} as implemented by \citet{barklem_kaulakys:_2016-1}. Again, these rates are significantly smaller than those estimated using the Drawin formula by 2--5 orders of magnitude. Very large deviations appear for the transitions with the lowest and highest rates, which the Drawin formula overestimates by as much as 10 orders of magnitude. \citet{barklem_inelastic_2011} conclude in their review that this poor agreement is due to the fact that low-energy inelastic hydrogen collisions are quantum mechanical in character, and these physics simply are not considered in the Drawin formula.

We adopt excitation rates due to electron collisions from open-ADAS\footnote{ADF04 data products for \ion{Al}i and \ion{Al}{ii} computed in 2012, available online at \url{http//open.adas.ac.uk}} \citep[computed as described by][]{badnell2011} when possible, including forbidden transitions.
These data are limited to states of \ion{Al}i with $n \le 4$, but cover all states of \ion{Al}{ii} included in our model atom.
For the remaining allowed \ion{Al}i transitions, we compute rates following \citet{seaton_impact_1962}.
In the literature, rates of forbidden transitions are typically estimated using \citet{vanregemorter1962} and \citet{bely1970}, and adopting some constant strength parameter.
We instead follow an approach similar to that of \citet{osorio_mg_2015-1} by examining the collision strength $\Upsilon_{ij}/g_i$, separately for spin-exchange and non-exchange transitions. We find that at a given temperature, forbidden non-exchange transition strengths correlate well with the transition energy, while spin-exchange transitions do not. Hence, we utilize a linear fit to $\log ((E_j-E_i) / \text{eV})$--$\log \Upsilon_{ij}/g_j$ for the remaining forbidden non-exchange transitions. Transition strengths for spin-exchange transitions vary at a given temperature by two orders of magnitude with no obvious correlation, so we simply adopt their mean value.
For electron collisional ionisation, we use the empirical formulae given by \citet[Sect.~3.6.1]{cox_allens_2000}, which originate from \citet{percival_cross_1966}.

\subsection{Spectrum analysis}
We use the spectrum synthesis code SME \citep{valenti_spectroscopy_1996,piskunov_spectroscopy_2017}, which allows parameter optimization while applying departure coefficients interpolated from grids which we precompute as a function of $\Teff$, $\logg$, $\FeH$ and $\eH{Al}$. 
\citet{piskunov_spectroscopy_2017} show that their heuristic uncertainty estimation based on the distribution of pixel-by-pixel errors produces realistic error bars, in comparison to $\chi^2$-based errors which typically strongly underestimate the uncertainties. We note that the former method breaks down when the number of pixels is very small, and adopt the larger of the two estimates.

The SME code is run using a custom pipeline in an unattended mode, using predefined windows for continuum normalization and line profile fitting. 
Background line data is taken from VALD3 \citep{ryabchikova_major_2015,piskunov_vald:_1995}, and we apply NLTE corrections to lines of Fe using the grid of \citet{lind_non-lte_2012}.
The adopted stellar parameters are listed in Table~\ref{tbl:stellarparams}.

For broadening due to collisions with neutral hydrogen, VALD and SME use a format where the broadening cross-section $\sigma$ and velocity parameter $\alpha$ are represented in a compressed format as $\text{int}(\sigma) + \alpha$.
For several transitions where the upper state is highly excited, the velocity parameter $\alpha$ is in fact greater than one. In these cases, we adopt $\alpha = 0.999$ in the abundance analysis and confirmed that this choice has no influence on the emergent line profiles.

Our 3D NLTE abundance analysis of the Sun is performed differentially with respect to the SME analysis in \avgtd NLTE, by fitting unblended 3D NLTE spectra to an unblended \avgtd NLTE reference synthesis. For consistency, we repeat the analysis in LTE to produce 3D LTE abundance estimates.

\subsection{Observational data}

We analyse the IAG atlas of the solar flux spectrum \citep{reiners_iag_2016}, as well as the disk-centre Li\`ege atlases recorded at Jungfraujoch \citep{delbouille_atlas_1973,delbouille_jungfraujoch_1995} and Kitt Peak \citep{delbouille_photometric_1981}.

We examine centre-to-limb variations using the $\mu \approx 0.15$ optical FTS atlas described by \citet{stenflo_coherent_1983}, which \citet{stenflo_fts_2015} paired with the Hamburg disk-centre FTS atlas \citep[made available online by][]{neckel_announcement_1999}. They recorded spectra near the solar limb using a $17.5" \times 10"$ aperture set parallel to the limb, thus covering a range $\mu \approx 0.10$--0.17, with an expected uncertainty of at least 0.01 in $\mu$ due to limitations in the guiding system. 
We also use high-resolution ($R \approx 150\,000$) spectra taken with the Swedish Solar Telescope \citep[SST][]{scharmer_1-meter_2003} using the TRIPPEL spectrograph \citep[][]{kiselman_is_2011}.
The data cover the 7835--7836\,\ang doublet at $\mu = 0.2$--1.0, and is further described by \citet{lind_non-lte_2017}.

We analyse Arcturus using the FTS atlases of \citet{hinkle_visible_2000} in the optical ($R \approx 150\,000$) and \citet{hinkle_infrared_1995} in the IR ($R \approx 100\,000$). Both atlases were recorded with a FTS \citep{hall_1.4_1979} at KPNO.

We also analyse the standard stars Pollux, \BSTOP, \BSSGB and \BSRGB using $R \approx 80\,000$ spectra from UVES-POP \citep{bagnulo_uves_2003}. 
For \BSMS, we use a $R \approx 70\,000$ NARVAL spectrum obtained via Polarbase \citep{petit_polarbase:_2014}.
In the UV, we use a high-resolution ($R \approx 114\,000$) HST-STIS spectrum of \BSSGB in the region 2100--3100\,\ang, and medium-resolution ($R \approx 30\,000$) spectra of \BSTOP and \BSRGB in the region 2300--3100\,\ang. The reduced spectra come from the ASTRAL catalogue \citep{ayres_advanced_2013}. 
In the IR, we use a medium-resolution ($R \approx 22\,500$) APOGEE \citep{majewski_apache_2016} spectrum of \BSRGB. 
For Pollux, we use two high-resolution ($R > 120\,000$) FTS spectra from KPNO recorded in September 1979 and April 1993 (K. Hinkle, priv. comm.).

In the mid-IR, we use the spectra at 12.33\,\micron from \citet{sundqvist_mg_2008} for Arcturus and Pollux, and the center-to-limb solar spectra from \citet[via J. Sundqvist, priv. comm.]{brault_solar_1983}.



\section{Results}\label{sect:results}
We present a detailed analysis of the Sun in Sect.~\ref{sect:sun}.
These include full 3D NLTE spectrum synthesis and analyses of center-to-limb variations (Sect.~\ref{sect:solar_profiles}). 
These are the very first 3D NLTE calculations for solar aluminium lines, as well as the very first center-to-limb analyses of optical aluminium lines.
Our solar abundance analysis indicates a 3D NLTE value of $\Abund{Al} = 6.43 \pm 0.03$, in exact agreement with the meteoritic value 
$\Abund{Al} = 6.43 \pm 0.01$ (\citealt{lodders_abundances_2009}, renormalized to the abundance of Si derived by \citealt{scott_elemental_2015-1} and \citealt{amarsi_solar_2017}). 

In Sect.~\ref{sect:kgiants}, we perform comprehensive analyses of the K-giants Arcturus and Pollux, including the mid-IR emission lines, and find good agreement with modeling.
Finally, we present in Sect.~\ref{sect:metalpoor} an abundance analysis of four metal-poor stars, which for the first time includes aluminium lines in the near-UV as well as the near-IR, again indicating very good agreement in the abundances derived from different aluminium lines.
Stellar parameters and abundance results are summarized in Table~\ref{tbl:stellarparams} for our benchmark stars, along with previous results for the ultra-metal poor red giant SMSS 0313-6708 \citep[from][]{nordlander_3d_2017}.

\begin{table*}
\caption{Stellar parameters of standard stars, and results of the abundance analyses.}
\label{tbl:stellarparams}
\centering
\begin{tabular}{l l l r c c c c ccccc c}
\hline\hline \noalign{\smallskip}
Star     & $\Teff$   & $\logg$      & $\FeH$    & $\vmic$ & Reference     & $\Abund{Al}$    & $\Abund{Al}$    & $\Abund{Al}$    & $\Abund{Al}$    \\
         &$[\text K]$&$[\text{cgs}]$&           & $[\kms]$&               &    1D LTE       &   1D NLTE       &   \avgtd NLTE   &   3D NLTE       \\ 
\hline \noalign{\smallskip}
Sun      &    5771   &    4.44      &     0.00  &  1.09   & B12, H15      & $6.42 \pm 0.03$ & $6.42 \pm 0.03$ & $6.46 \pm 0.03$ & $6.43 \pm 0.03$ \\ 
Arcturus &    4247   &    1.59      &   $-0.52$ &  1.63   & J14,      O15 & $6.23 \pm 0.08$ & $6.11 \pm 0.05$ & $6.10 \pm 0.05$ \\ 
Pollux   &    4858   &    2.90      &     0.13  &  1.28   & J14, H15, J15 & $6.55 \pm 0.04$ & $6.55 \pm 0.04$ & $6.54 \pm 0.06$ \\ 
\BSMS    &    5786   &    4.23      &   $-0.86$ &  1.16   & J14           & $5.72 \pm 0.20$ & $5.78 \pm 0.08$ & $5.79 \pm 0.08$ \\ 
\BSTOP   &    6408   &    4.13      &   $-2.03$ &  1.40   & B12           & $3.47 \pm 0.15$ & $3.85 \pm 0.14$ & $3.87 \pm 0.12$ \\ 
\BSSGB   &    5777   &    3.67      &   $-2.40$ &  1.23   & B12, J14, O15 & $3.14 \pm 0.14$ & $3.57 \pm 0.12$ & $3.52 \pm 0.12$ \\ 
\BSRGB   &    4608   &    1.61      &   $-2.64$ &  1.50   &      J14, O15 & $3.56 \pm 0.25$ & $3.65 \pm 0.17$ & $3.65 \pm 0.10$ \\ 
SMSS0313\tablefootmark a
         &    5125   &    2.30      & $< -6.53$ &  2.00   & K14, N17      & $< 0.74$        & $<1.29$         &  $< 1.22$       &  $ < 1.30$ \\ 
\hline
\end{tabular}
\tablefoot{
The abundance results are unweighted arithmetic mean values, with the uncertainty signifying the line-to-line scatter, except for the Sun where lines are weighted and systematic errors are included in the uncertainty. 
The recommended abundance is given in the rightmost available column for every star.
\tablefoottext a {Upper limits ($3 \sigma$) on Al from \citet{nordlander_3d_2017}.}
}
\tablebib{
B12: \citet{bergemann_non-lte_2012};
J14: \citet{jofre_gaia_2014}; 
K14: \citet{keller_single_2014}; 
H15: \citet{heiter_gaia_2015}; 
J15: \citet{jofre_gaia_2015}; 
O15: \citet{osorio_mg_2015-1}; 
N17: \citet{nordlander_3d_2017}.
}
\end{table*}

\subsection{Solar analysis} \label{sect:sun}

\begin{figure}
	\includegraphics[width=.47\textwidth]{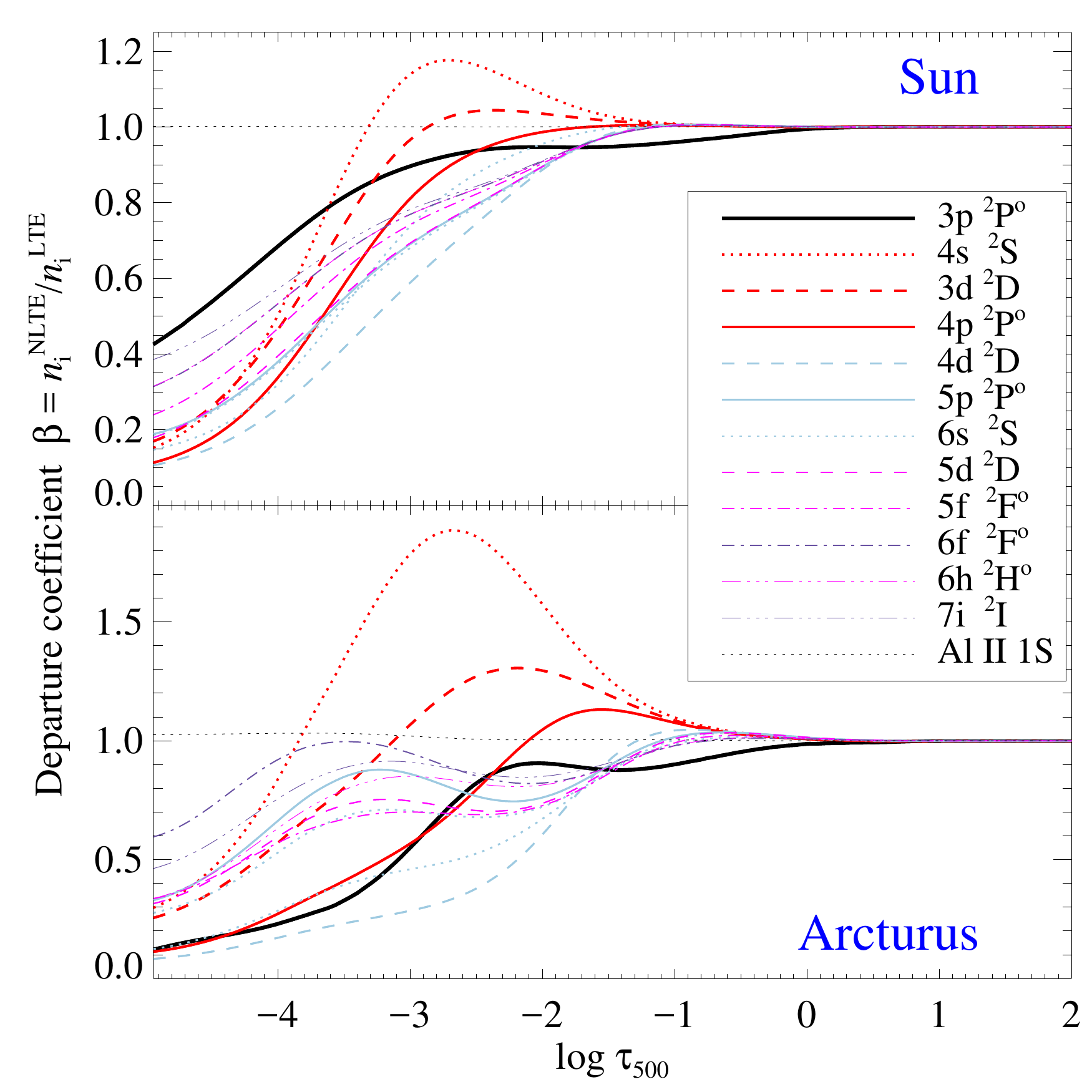}
\caption{Departure coefficients in \avgtd models representing the Sun and Arcturus, computed at abundances representative of those found in the abundance analysis.}
\label{fig:depcoeffs}
\end{figure}

Departure coefficients computed for an \avgtd \stagger model representing the Sun are shown in Fig.~\ref{fig:depcoeffs}. 
The NLTE mechanisms in a solar atmospheric model have previously been described in detail by \citet{baumuller_line_1996}, and are largely in agreement with our results. 
Three main radiative mechanisms are at play, and their detailed interplay determine the net effects.

First, photoionization of the ground $3 \rm p$ state causes it to deplete relative to the LTE population as deep into the solar photosphere as $\log \tau_{500} = 0$. 
In general, this effect strengthens with increasing effective temperature and decreasing metallicity, both of which lead to an increasingly superthermal ultraviolet radiation field.
Photoionization also depletes the $3 \rm d$ and $4 \rm p$ states outside $\log \tau_{500} = -3$ in the solar photosphere, where the decreasing pressure leads to relatively inefficient collisional couplings.

Second, as the resonance lines have very large radiative transition probability, their source function $S_\nu^l$ is coupled to the mean radiation field $\bar J_\nu$. 
At small continuum optical depths, $\log \tau_{500} < -3$, the line core is optically thick, causing the mean radiation field to drop below the Planck function in a process called resonance scattering. Since $\beta_\text j / \beta_\text i \propto S^l_\nu = \bar J_\nu$, the upper state $4 \rm s$ is depopulated relative to the ground state due to photon losses. 
At lower metallicity, the lines weaken and thus the resonance scattering effect disappears. While the source function is still coupled to the mean radiation field, the superthermal radiation field instead drives photon pumping, which overpopulates the upper state $4 \rm s$ and even more highly excited $n \rm s$ and $n \rm d$ states.

Third, a cascade of close-lying infrared transitions with $\Delta l = -1$ (thick lines in Fig.~\ref{fig:termdiag}) participate in a process called ``photon suction'' \citep{bruls_formation_1992}. For each such transition, the subthermal radiation field causes photon losses which drives a downward flow.
The cascade of such processes causes relative overpopulations in the low-excitation $4 \rm s$, $3 \rm d$ and $4 \rm p$ states, while depleting the more highly excited states.

In addition to these radiative processes, close-lying energy levels are coupled mainly by hydrogen collisions for low-excitation states and by electron collisions for highly excited states.
The $3 \rm d$ and $4 \rm p$ states are efficiently coupled via hydrogen charge transfer to the \ion{Al}{ii} continuum which follows LTE. 
Highly excited Rydberg states are coupled by electron collisions in a cascade upwards to the \ion{Al}{ii} continuum. As the $\rm 4 d$ state is strongly depopulated (due to photon suction), the collisional couplings cause the more highly excited states to become successively less underpopulated. This results in an inversion of populations among these states, where $\beta_u / \beta_l > 1$ results in superthermal source functions (which are enhanced due to stimulated emission) for permitted transitions among the Rydberg states.

\subsubsection{The solar 3961\,\ang resonance line}
\label{sect:solar_profiles}

\begin{figure}
	\includegraphics[width=.47\textwidth]{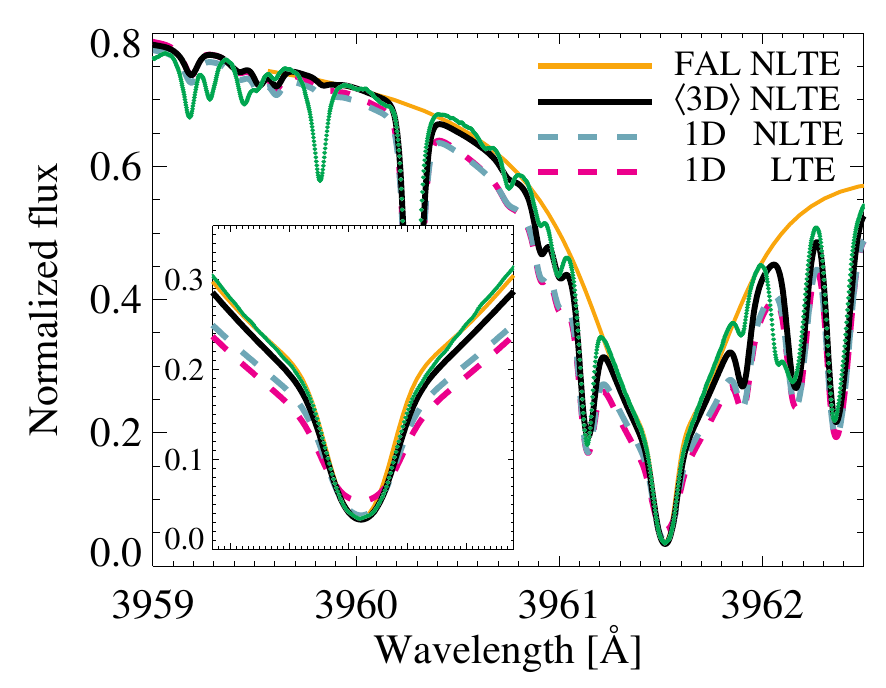}
\caption{Solar resonance line, comparing the KPNO flux atlas to NLTE synthesis using the FALC chromospheric solar model, as well as \stagger \avgtd and \marcs 1D models, computed with $\Abund{Al} = 6.43$. LTE synthesis is also shown for the 1D \marcs model. The FALC spectrum does not take into account minor line blends. The inset shows a zoom on the core region.}
\label{fig:solar_resline}
\end{figure}

We illustrate the \ion{Al}i resonance line at 3961\,\ang in Fig.~\ref{fig:solar_resline}, and find that it is well reproduced by our NLTE modeling. In particular, the flux level in the core matches observations to within 1\,\% under NLTE using both 1D and \avgtd models. 
The FALC model shows an equally good fit, implying a negligible contribution from the chromosphere.
Applying LTE synthesis to the FALC model (not shown), the chromospheric temperature inversion would produce unobserved core emission on the level of 25\,\% of the continuum flux. In NLTE however, the ground state is strongly overpopulated in the chromospheric layers, resulting in a subthermal source function fainter than the photospheric continuum at $\log \tau_{500} = -3$, which fully suppresses emission. 

In the photospheric regions, the behaviour is similar in all three model atmospheres:
The upper state of the transition ($4 \rm s$) is more strongly depleted than the ground state at the very small optical depths, $\log \tau_{500} < -3$, where the line core forms, causing the source function to become subthermal and leading to a darkened, deeper core. At optical depths between $\log \tau_{500} = 0$ and $-3$, where the inner wings form, the $4\rm s$ state is slightly overpopulated relative to LTE. This causes a brightening of the source function, thus weakening the wings. 
As the 3961\,\ang line is located well within the blue wing of the \ion{Ca}{ii} H 3968\,\ang line, the former is significantly depressed by the latter.
The \ion{Al}i line is thus well reproduced with both 1D and \avgtd models, if the Ca abundance is fine-tuned.

\subsubsection{Center-to-limb variations in the 7835\,\ang line}

\newcommand\bcut{1.9} 
\begin{figure*}
\centerline{
	\includegraphics[clip,trim=0 \bcut em 0 0]{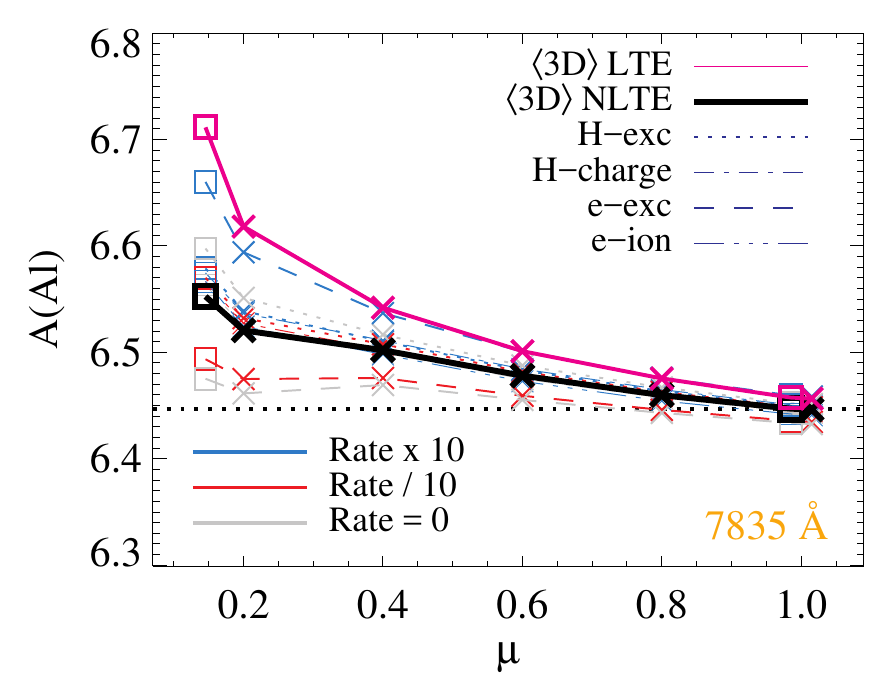}
	\includegraphics[clip,trim=0 \bcut em 0 0]{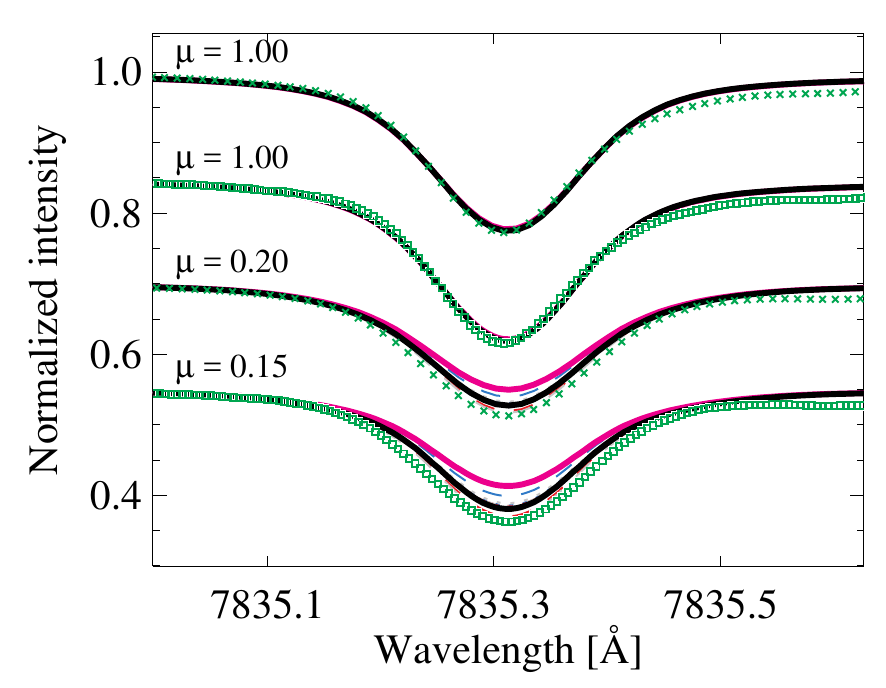}
}
\centerline{
	\includegraphics[clip,trim=0 \bcut em 0 0]{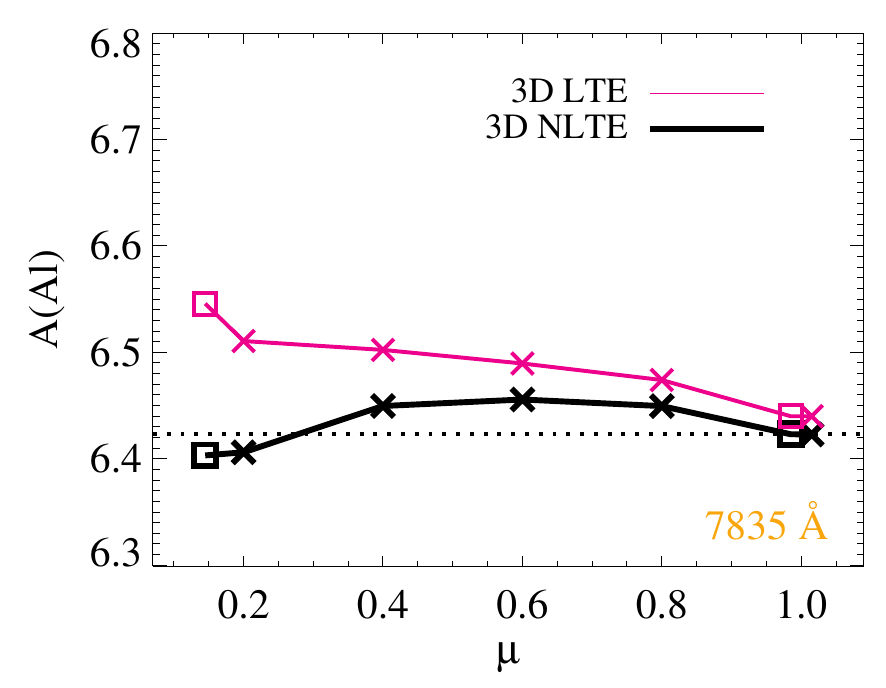}
	\includegraphics[clip,trim=0 \bcut em 0 0]{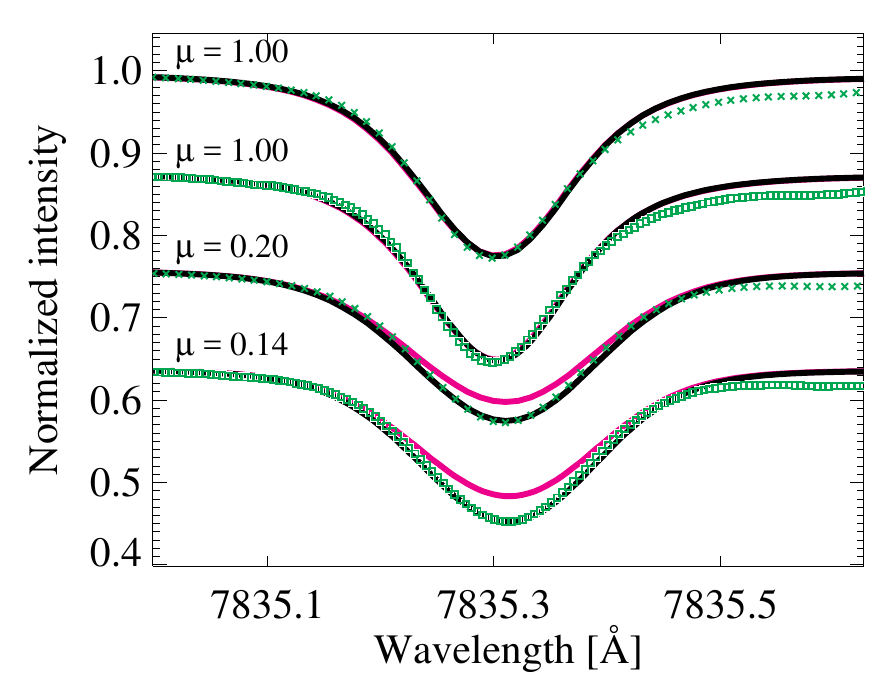}
}
\caption{Center-to-limb variations shown for the solar spectrum computed in LTE and NLTE using \avgtd (above) or 3D models (below), illustrated as the fitted abundances (left) and the predicted line profiles (right). Observations come from the FTS CLV atlas (squares) and TRIPPEL (greek crosses). For the \avgtd model, rates describing electron collisional excitation (e-exc) and ionization (e-ion), and hydrogen collisional excitation (H-exc) and charge transfer (H-charge) have been increased or decreased by a factor 10 (green and red), or removed completely (gray). The LTE modeling (solid magenta line) always leads to the \textit{highest} abundances, and \textit{increased} and \textit{decreased} collisional rates give higher or lower abundances than the standard NLTE model, respectively. }
\label{fig:clv}
\end{figure*}

We illustrate the center-to-limb variation for the 7835\,\ang line of \ion{Al}i in Fig.~\ref{fig:clv}. The line profiles illustrate the predicted variation in line strength toward the limb. As reference, we use the abundance determined at $\mu = 1$, and note that we find the same result for the FTS and TRIPPEL spectra to within 0.001\,dex. Additionally, we find exceedingly small NLTE effects at disk center (in fact for all optical lines of \ion{Al}i), with abundance corrections of less than 0.01\,dex for \avgtd models.
The average formation depth at disk center is greater than $\log \tau_{500} = -1.5$, where all relevant populations exhibit near-LTE populations. Toward the limb, line formation shifts toward higher layers, where deviations from LTE become apparent.
While the lower state $3 \rm d$ is coupled to the \ion{Al}{ii} continuum via charge transfer and thus has an LTE population, the upper state $6 \rm f$ is depleted due to photon losses in the line. This results in a subthermal source function, which darkens and thus strengthens the line core.

We find that 3D NLTE synthesis successfully predicts the CLV to within 0.03\,dex, which is comparable to expected errors due to uncertainties in the continuum placement and unidentified blends. 
Minor uncertainties in positioning on the solar disk have similar impact, as a shift in $\mu$ by 0.02 near the limb would affect the inferred abundances by 0.02\,dex.
In contrast to our 3D NLTE synthesis, \avgtd models underestimate the line strength toward the limb, corresponding to an error in abundance of $+0.10$\,dex at $\mu = 0.15$.

For our \avgtd NLTE modeling, we have modified transition rates due to collisions with hydrogen atoms and with electrons by multiplying or dividing by a factor 10. 
The line strength is most sensitive to electron collisional excitation, where an increase or decrease in rates by a factor 10 respectively affects the inferred abundance by $+0.09$ or $-0.05$\,dex at $\mu = 0.15$, and by smaller amounts (with the same sign) at larger viewing angles. 
In contrast, modifying rates for hydrogen collisional excitation, charge transfer or electron collision ionization up or down by a factor 10 increases the inferred abundance by 0.02--0.03\,dex at $\mu = 0.15$, and by smaller positive amounts at larger viewing angles.
The very good agreement of our 3D NLTE line profiles with observations thus imply that errors in electron collisional rates are likely significantly smaller than a factor 10.
If our modeling like previous studies had been limited to 1D or \avgtd modeling, the mismatch between 1D or \avgtd NLTE modeling and observations could have been misinterpreted as a consequence of erroneous collisional rates, rather than a shortcoming of the atmospheric model.

\subsubsection{The 12.33\,\micron emission line}

\begin{figure*}
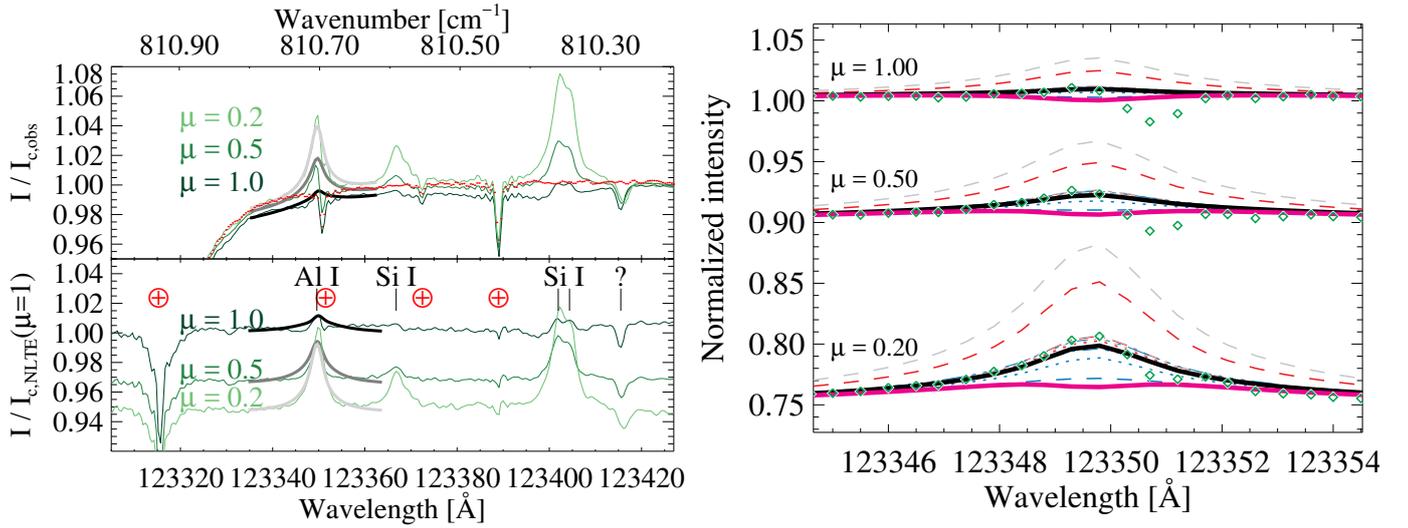

\centerline{
	\includegraphics[]{{{figures/irlines_m3d}}}
	\includegraphics[]{{{figures/profiles_solar-CLV-12.335-stagger-collisions}}}
}
\caption{Emission line profiles in the mid-IR, comparing solar intensity spectra to 3D NLTE synthesis (left) and \avgtd NLTE and LTE synthesis (right) at different disk positions.
\textit{Left panel}: intensity spectra surrounding the 12.33\,\micron emission line. Positions toward the limb are shown in fainter colors, and thick lines represent 3D NLTE synthesis. Tellurics are indicated in the lower panel, while stellar lines are labeled. 
In the \textit{top panel}, observed spectra have been normalized according to continuum regions outside the plotted region, to highlight center-to-limb variations in the local continuum. The synthetic spectra have been normalized to the observed local continuum, including telluric absorption, and a template telluric spectrum is shown (red dots). 
In the \textit{bottom panel}, observations have been divided by the telluric spectrum and then renormalized relative to the predicted continuum intensity at disk centre, to show the predicted amount of limb darkening in the continuum.
\textit{Right panel}: Zoom of normalized intensity spectra near the \ion{Al}i emission line, arbitrarily offset for different disk positions, without telluric corrections.
Synthetic spectra are shown computed in \avgtd LTE (solid magenta), \avgtd NLTE (solid black), and in NLTE with different modifications to collisional transition rates, as in Fig.~\ref{fig:clv}.}
\label{fig:MIRsun}
\end{figure*}

Highly NLTE-sensitive lines due to transitions between Rydberg levels are found in the mid-IR. In particular, the emission lines found in the solar spectrum at 12\,\micron \citep{murcray_observation_1981} have been convincingly shown to be due to \ion{Mg}i, \ion{Al}i and \ion{Si}i \citep{chang_identification_1983,chang_non-penetrating_1984}.
Observations at and just above the limb indicated an origin in either the high photosphere or very low chromosphere \citep{brault_solar_1983}, with further theoretical study indicating a photospheric origin, with the emission caused by a population inversion \citep{chang_formation_1991,carlsson_formation_1992}. 
As the emission line strength is sensitive to collisional and radiative couplings among the highly excited states, the Mg lines have since been studied at length in the literature \citep[\eg][]{ryde_zeeman-sensitive_2004,sundqvist_mg_2008,osorio_mg_2015-1}. 
The lines of \ion{Al}i have gathered less interest, as the formation mechanism is essentially the same. Nonetheless, \citet{baumuller_line_1996} were able to reproduce the emission line profiles of the lines at 11.93 and 12.26\,\micron, after finetuning their hydrogen collision transition rates. 

We show our predicted center-to-limb variations of the $6\rm h$--$7\rm i$ transition at 12.33\,\micron in Fig.~\ref{fig:MIRsun}. 
The line is rather faint, and observations are essentially flat at disk center with an emission peak of 6\,\% at $\mu = 0.20$. A telluric absorption line is present at 123350.5\,\ang, causing a 2\,\% dip in the line profile, and the strong telluric line at 123315\,\ang strongly perturbs the continuum. 
We show results of both 3D NLTE synthesis and \avgtd NLTE and LTE synthesis, but note that the normalized line profiles are essentially identical. 
The synthetic line profile appears to be weaker than observations, with emission at the limb of 5\,\%, possibly indicating insufficient population reversal in our modeling.
The strength of this feature is most sensitive to the electron collisional transition rates, and we find that small changes (factor of two) would significantly affect the line strength. 
In contrast, the sensitivity to hydrogen collisional rates is very small, and an increase by a factor ten in these rates would only slightly weaken the line.

We have performed NLTE calculations in full 3D NLTE, as well as using \avgtd, \marcs and \falc \citep{fontenla_energy_1993} model atmospheres.
The normalized line profiles at $\mu = 0.2$ are essentially indistinguishable, with emission peaks varying in strength from 4.8\,\% to 5.2\,\%, indicating that results are not significantly sensitive to small differences in the photospheric temperature structure or horizontal inhomogeneities. 
In the \falc atmosphere, the line still forms well above the temperature minimum, and is thus not sensitive to the presence of a chromosphere. 

Analysis of lines in this region is however not straightforward. In the top left panel of Fig.~\ref{fig:MIRsun}, the local continuum is found to be depressed by 1\,\% at disk centre relative to the limb. We also find corresponding center-to-limb variations of 1\,\% in the continuum level surrounding the \ion{Mg}i 12.32\,\micron line, but not in the emission free region at 12.15\,\micron, indicating that it is not due to varying amounts of telluric absorption. It is thus not clear if this systematic behavior is caused by the continuum normalization procedure, or is intrinsic to the stellar spectrum due to extended absorption wings.

\subsubsection{Near-IR lines at 1.31 and 1.67\,\micron}

\begin{figure*}
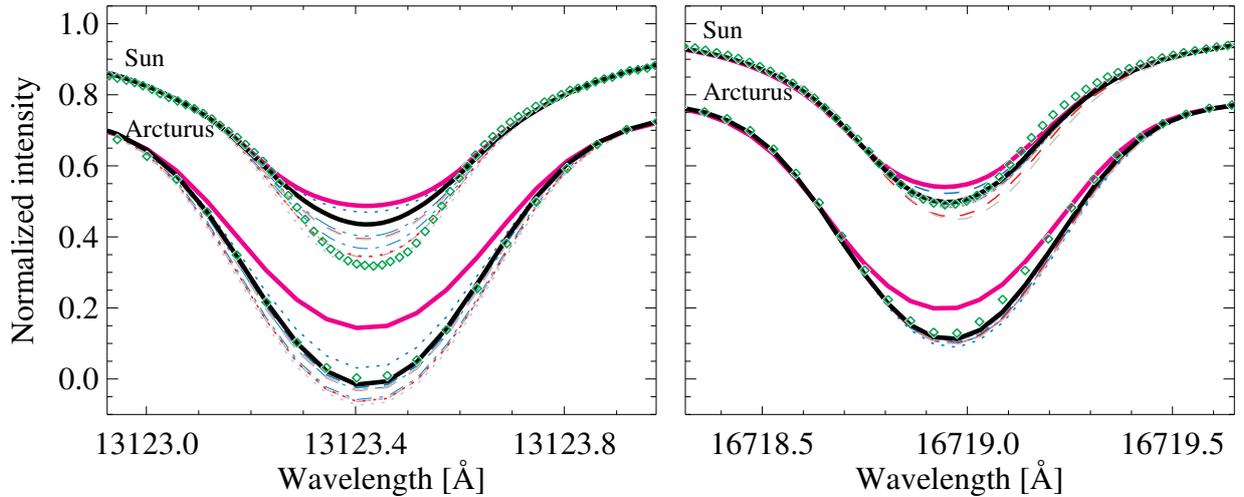

\centerline{
	\includegraphics[]{{{figures/profiles_sun-arcturus-13123-stagger-collisions}}}
	\includegraphics[clip, trim=4.2em 0 0 0]{{{figures/profiles_sun-arcturus-16718-stagger-collisions}}}
}
\caption{Spectra of two NLTE sensitive near-IR lines, comparing the Sun (at disk center) and the K-giant Arcturus, to synthetic \avgtd LTE (solid magenta) and \avgtd NLTE (solid black) profiles. The spectrum of Arcturus has been shifted downward by 0.2\,units. Additional NLTE profiles are shown where collisional transition rates have been modified, as in Fig.~\ref{fig:clv}. The synthetic profiles were computed assuming $\Abund{Al} = 6.43$ for the Sun, and $\Abund{Al} = 6.20$ for Arcturus}
\label{fig:NIRlines}
\end{figure*}

\begin{figure*}
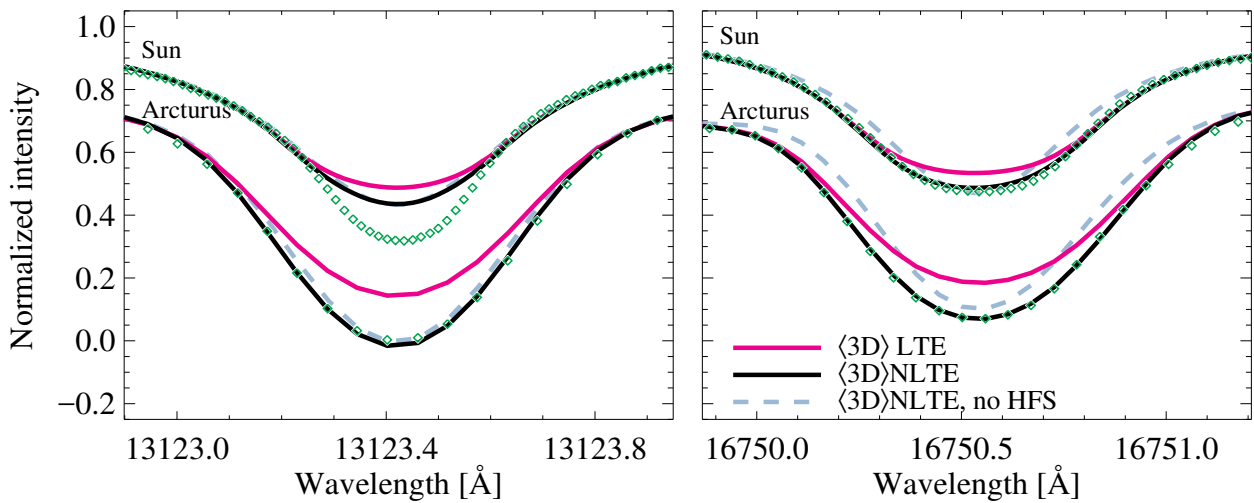

\centerline{
	\includegraphics[]{{{figures/profiles_hfs-sun-arcturus-13123-stagger-collisions}}}
	\includegraphics[clip, trim=4.2em 0 0 0]{{{figures/profiles_hfs-sun-arcturus-16750-stagger-collisions}}}
}
\caption{Spectra of two near-IR lines sensitive to hyperfine splitting, comparing the Sun (at disk center) and the K-giant Arcturus, to synthetic \avgtd LTE (solid magenta) and \avgtd NLTE (solid black) profiles. The spectrum of Arcturus has been shifted downward by 0.2\,units. An additional NLTE profile is shown where hyperfine splitting has been neglected (dashed blue). The synthetic profiles were computed assuming $\Abund{Al} = 6.43$ at 13123\,\AA\  and $6.33$ at 16750\,\AA\ for the Sun, and assuming $\Abund{Al} = 6.10$ and $5.95$ for Arcturus.}
\label{fig:HFSlines}
\end{figure*}

Less extreme examples of NLTE-sensitive lines are found in the near-IR, as illustrated in Fig.~\ref{fig:NIRlines}. 
The $\rm 4s$--$\rm 4p$ transition at 13123\,\ang exhibits a strongly darkened core due to the lower state being overpopulated (increasing the opacity) while the upper state is underpopulated (lowering the source function). 
The observed line core is however significantly deeper than what our NLTE synthesis produces, and while we show only \avgtd NLTE synthesis we note that the 3D NLTE line profile is very similar.
The line source function is very sensitive to collisional transition rates, and we find that a reduction in hydrogen collisional rates by a factor of a few (less than 10) would in principle reproduce observations.

In contast, the $\rm 4p$--$\rm 4d$ transition at 16718\,\ang forms deeper in the photosphere and exhibits less core darkening, in very good agreement with observations. 
The predicted NLTE effect is rather robust to uncertainties in collisional rates, and is thus not a sensitive diagnostic of modeling errors.
Larger NLTE effects are found in the 16750\,\ang line of the same multiplet, but as shown in Fig.~\ref{fig:HFSlines}, the comparison is somewhat muddled by strong hyperfine splitting of the upper level of the transition. 
The inclusion of hyperfine splitting significantly improves the fit to this line, affecting the solar abundance determination by $-0.1$\,dex. 
Splitting in the lower level of the 13123--13150\,\ang doublet is less significant, with abundance effects of $-0.01$ to $-0.05$\,dex.
These effects of hyperfine splitting are thus similar in magnitude to the NLTE corrections.

In conclusion, we find it unlikely that errors in collisional transition rates are larger than a factor 10 for either electrons (as indicated by the lines at 7835\,\ang and 12.33\,\micron) or hydrogen atoms (as indicated by the line at 13123\,\ang).

\subsubsection{Abundance analysis}
\label{sect:solar_abundances}

\begin{figure}
\centerline{
	\includegraphics[clip,trim=0 0 0 0]{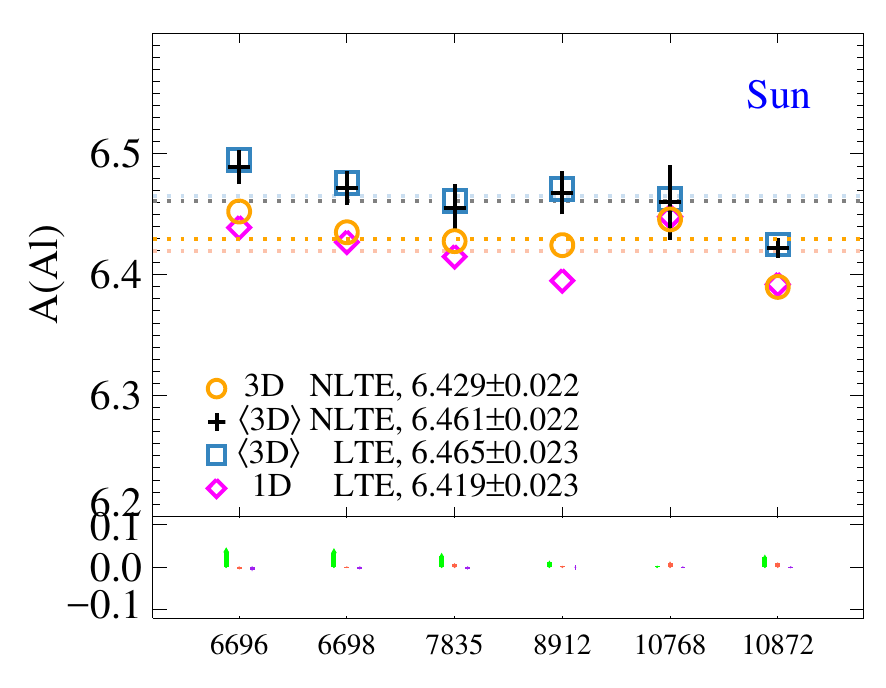}
}
\caption{Solar abundance analysis using the Li\`ege disk-center atlas. Abundances are shown for analyses using 3D NLTE (orange circles), as well as \avgtd models in both LTE (blue squares) and NLTE (black crosses), and 1D models in LTE (magenta diamonds). The average abundance from each analysis is indicated by the horizontal dashed lines. The height of the crosses indicate the statistical uncertainty in measuring the abundance from each line in the solar spectrum. 
Below, the sensitivity to changes in stellar parameters is shown for each line, indicating the effect of increasing $\Teff$ by 100\,K, $\logg$ by 0.3\,dex, and $\vmic$ by 0.3\,$\kms$. Adopting the weights from \citet{scott_elemental_2015-1} and including estimates of systematic uncertainties, the final 3D NLTE abundance is $\Abund{Al} = 6.43 \pm 0.03$.
}
\label{fig:solar_abund}
\end{figure}

Results of our disk-centre solar abundance analysis are illustrated in Fig.~\ref{fig:solar_abund}, where the average abundances are given as unweighted arithmetic mean values with uncertainties representing the line-to-line dispersion.
Adopting a line selection and weights from \citet{scott_elemental_2015-1}, but disregarding the line at 10891\,\AA\ due to telluric contamination, yields our recommended solar abundance, $\Abund{Al} = 6.43 \pm 0.03$, where the error takes into account systematic errors. 
This result is in perfect agreement with the meteoritic abundance, $\Abund{Al} = 6.43 \pm 0.01$ (\citealt{lodders_abundances_2009}, renormalized to the abundance of Si derived by \citealt{scott_elemental_2015-1} and \citealt{amarsi_solar_2017}).
We list results of our 3D NLTE analysis as well as the corresponding \avgtd NLTE, 1D NLTE and 1D LTE results in Table~\ref{tbl:stellarparams}. 
The fact that 3D NLTE and 1D LTE results are in very good agreement, is thanks to a near cancellation between effects of the on average steeper temperature structure in the hydrodynamical models ($\text{\avgtd} - \rm 1D \approx 0.04$\,dex) and horizontal inhomogeneities ($\rm 3D - \text{\avgtd} \approx -0.03$).
This cancellation should not be expected to occur in general, and in particular not when either NLTE effects or 3D effects are large. For example, the analysis of an ultra-metal poor red giant by \citet{nordlander_3d_2017} found a large 3D NLTE--1D LTE abundance difference of 0.6\,dex.

The NLTE abundance corrections adopted by \citet{scott_elemental_2015-1} are based on calculations using the atom of \citet{gehren_abundances_2004}, which in turn is an update from \citet{baumuller_line_1996}. The corrections were computed for flux spectra, and are thus somewhat larger than the corresponding disk-center corrections. However, their abundance corrections are \textit{positive}, while ours are negative. 
They determined a NLTE corrected 3D abundance of $\Abund{Al} = 6.43 \pm 0.04$, which is identical to ours as slight differences in the 3D model and line parameters happen to cancel with differences in the adopted NLTE corrections.

The uncertainty on our recommended solar abundance represents statistical and systematic errors combined in quadrature, following the approach of \citet{scott_elemental_2015-1}.
Here, we make the conservative assumption that all errors in collisional rates are smaller than a factor 10, which affects the average abundance by less than 0.01\,dex for disk-centre spectra, and note that the corresponding error for flux spectra is 0.02\,dex.


\subsection{K-giants: Arcturus and Pollux}
\label{sect:kgiants}

The departure coefficients computed for an \avgtd model representing Arcturus are illustrated in Fig.~\ref{fig:depcoeffs}. 
The departures from LTE are generally similar to those in the solar photosphere, as the same mechanisms are at play, but stronger due to lower concentrations of both hydrogen atoms and electrons resulting in weaker collisional couplings.
The ground $3 \rm p$ state is overionized by up to a factor 10 (resulting in a 1--2\,\% overpopulation of \ion{Al}{ii}), 
and the departure coefficients show a characteristic bump at $\log \tau_{500} = -2$ due to photon suction, which is more efficient than in the solar photosphere.
Despite the strong effect of overionization, photon suction dominates for the excited states which are responsible for the optical lines used in our abundance analysis.
For highly excited Rydberg states, the departure from LTE is stronger than in the Sun, resulting in a stronger divergence between departure coefficients and thus a stronger emission mechanism for the 12\,\micron line.

\subsubsection{Line profile comparisons}

\begin{figure}
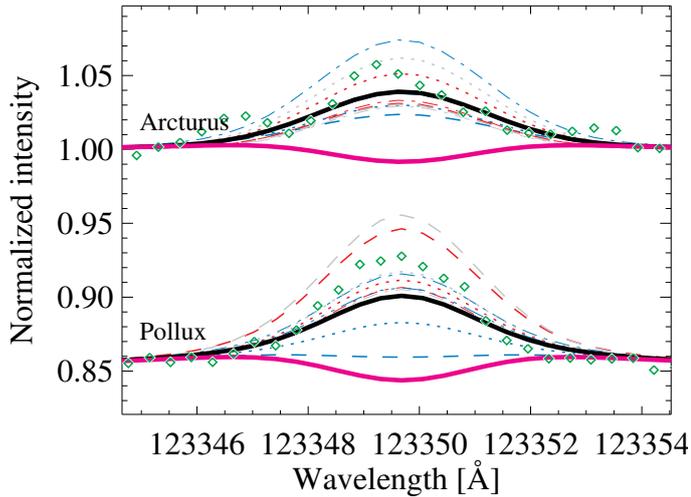

\centerline{
	\includegraphics[]{{{figures/profiles_Kgiant-12.335-stagger-collisions}}}
}
\caption{Emission line profiles in the mid-IR, comparing the K-giants Arcturus and Pollux, to synthetic \avgtd LTE (solid magenta) and \avgtd NLTE (solid black) profiles computed using \multi. The continuum level has been arbitrarily offset for Pollux. Line styles and colors have the same meaning as in Fig.~\ref{fig:MIRsun}.
The synthetic profiles were computed assuming $\Abund{Al} = 6.20$ for Arcturus, and $\Abund{Al} = 6.65$ for Pollux.}
\label{fig:MIRlines}
\end{figure}

As shown in Fig.~\ref{fig:MIRlines}, the observed emission line in the spectra of Arcturus and Pollux are rather similar, with an emission peak of 6--7\,\%.
In the spectrum of Arcturus, our NLTE modeling is in good agreement with observations, relative to the spectrum noise level. 
For Pollux, which has 1.3\,dex higher surface gravity and thus significantly higher collisional rates, our NLTE model predicts a somewhat weaker than observed emission line. Decreasing the collisional rates, in particular electron collisional rates, by a factor of a few would improve the fit for both stars.
The sensitivity of this modeling to the adopted model atmosphere is small for both stars, as NLTE synthesis using \marcs models predicts emission cores weaker by 1--2\,\% relative to the continuum.

The near-IR lines of Arcturus, illustrated in Fig.~\ref{fig:NIRlines} are in excellent agreement with observations.
The core darkening which is sensitive to NLTE effects on the $\rm 4 s$, $\rm 4 p$ and $\rm 4 d$ states is well reproduced, but also appears to be rather robust in regard to uncertainties in the collisional transition rates.

\subsubsection{Abundance analysis}

\begin{figure*}
\centerline{
	\includegraphics[clip,trim=0 0 0 0]{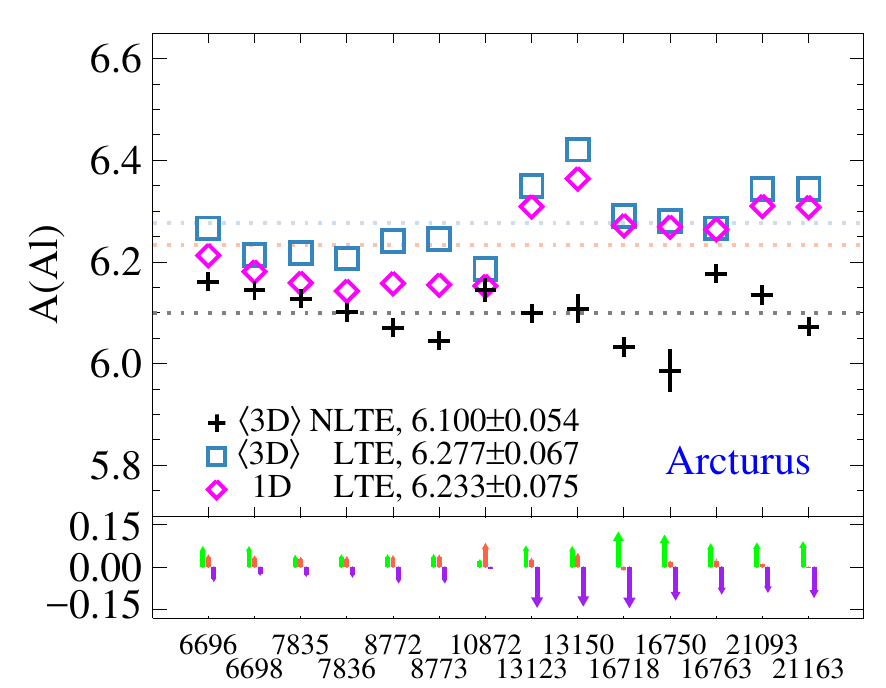}
	\includegraphics[clip,trim=\bcut em 0 0 0]{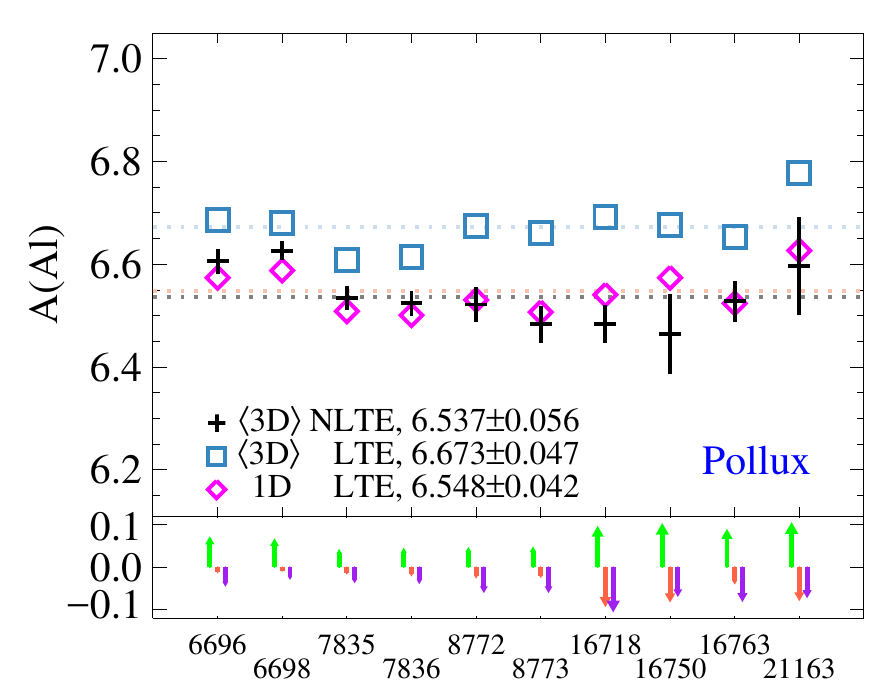}
}
\caption{Abundance analysis of Arcturus using spectra from KPNO, and Pollux using spectra from UVES-POP (optical) and KPNO (infrared).}
\label{fig:abundances_arcturus}
\end{figure*}

We illustrate our abundance analysis of Arcturus in Fig.~\ref{fig:abundances_arcturus}.
Our analysis of 15 lines in the optical and near-IR indicate that NLTE and \avgtd--1D effects are typically both significant, but with opposite sign.
NLTE modeling always yields lower abundances, and this effect ranges in magnitude between 0.04\,dex for the very weak line at 10872\,\ang, and 0.31\,dex for the strongly saturated 13150\,\ang line. 
The NLTE effect is stronger in the \avgtd models, and we find that 1D NLTE and \avgtd NLTE analyses agree to within 0.03\,dex for every line. 
We find somewhat smaller line-to-line scatter using our \avgtd NLTE modeling than in LTE, indicating an arithmetic mean abundance $\Abund{Al} = 6.10 \pm 0.05$, meaning $\eH{Al} = -0.33$ or $\eFe {Al} = 0.19$. 
The near-IR lines are often strongly saturated, making them sensitive to the stellar parameters. Due to their strength, we also expect larger systematic uncertainties due to missing or erroneous blending lines, which propagates into both continuum placement and the line profile itself. 
Notably, the strongly saturated line at 16750\,\ang is significantly broadened by hyperfine splitting, and we do not recommend this line for abundance analyses of this type of star unless both NLTE effects and hyperfine splitting are taken into account.

For Pollux, we find that while NLTE effects range between $-0.05$ and $-0.21$\,dex, these corrections and the \avgtd--1D effects on average nearly cancel for both optical and infrared lines. The arithmetic mean abundance $\Abund{Al} = 6.54 \pm 0.06$ has a scatter comparable to that found for Arcturus.

\subsection{Metal-poor stars}
\label{sect:metalpoor}

In very metal-poor stars with $\FeH < -2$, the only suitable aluminium line is the resonance line at 3961\,\ang. 
While the optical lines become accessible at $\FeH > -1$, the resonance line at the same time becomes unsuitable for abundance analysis due to severe saturation and blending with the nearby \ion{Ca}{ii} H 3968\,\ang line. 
This introduces a bimodality in literature studies, where either the resonance line or the optical lines are used.

\begin{figure*}
	\includegraphics[width=\textwidth]{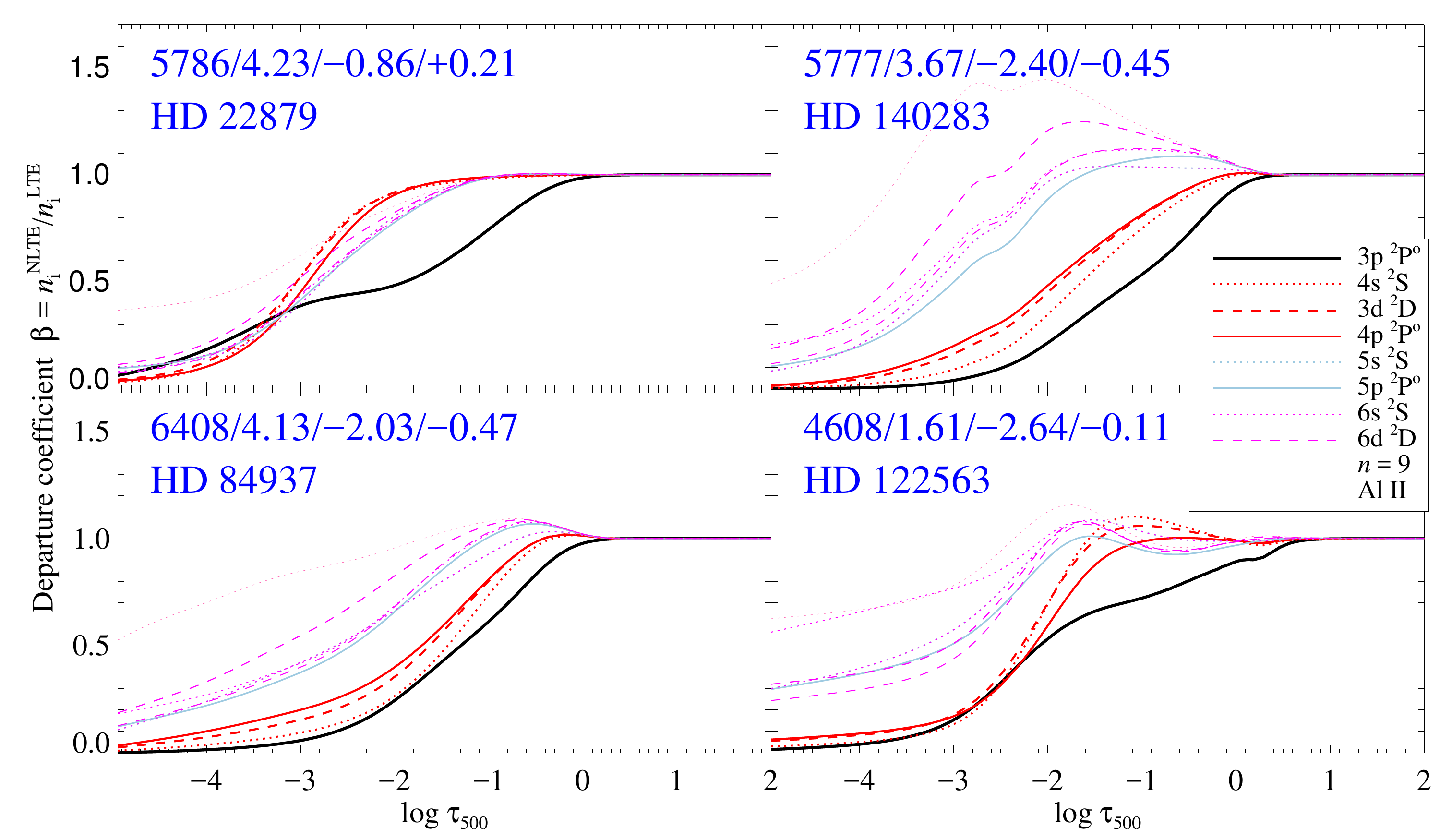}
\caption{Departure coefficients in \avgtd models representing four metal-poor stars. The parameters of the model, $\Teff/\logg/\FeH/\eFe{Al}$, are indicated in each panel.}
\label{fig:depcoeffs2}
\end{figure*}

\begin{figure*}
\centerline{
	\includegraphics[clip,trim=0 0 0 0]{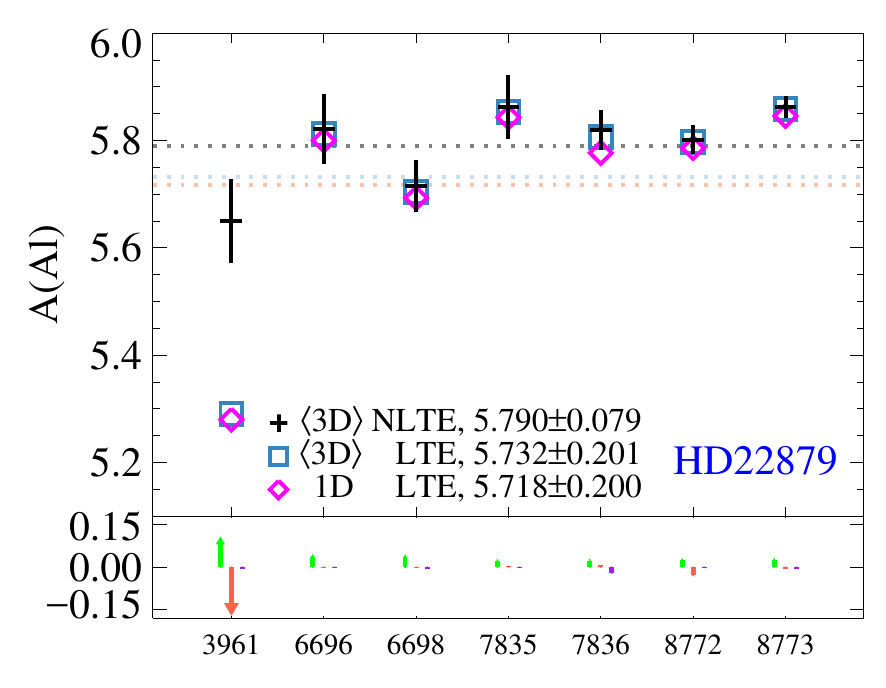}
	\includegraphics[clip,trim=\bcut em 0 0 0]{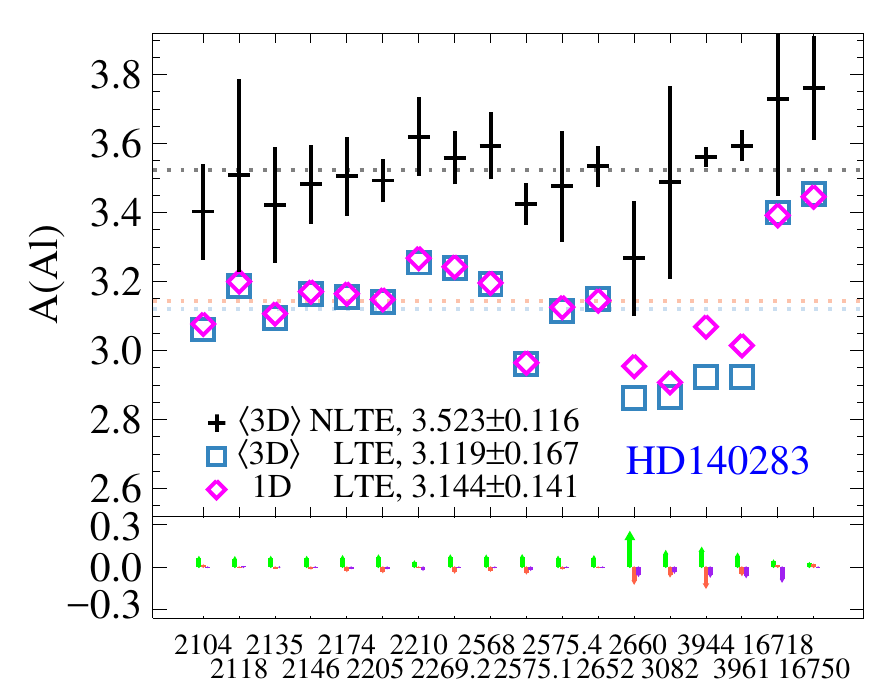}
}
\centerline{
	\includegraphics[clip,trim=0 0 0 0]{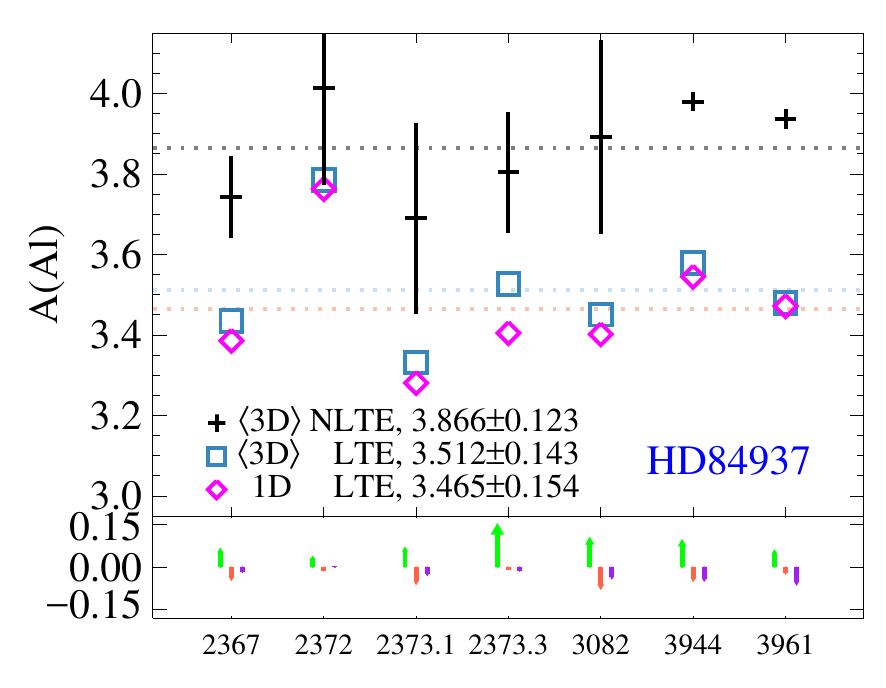}
	\includegraphics[clip,trim=\bcut em 0 0 0]{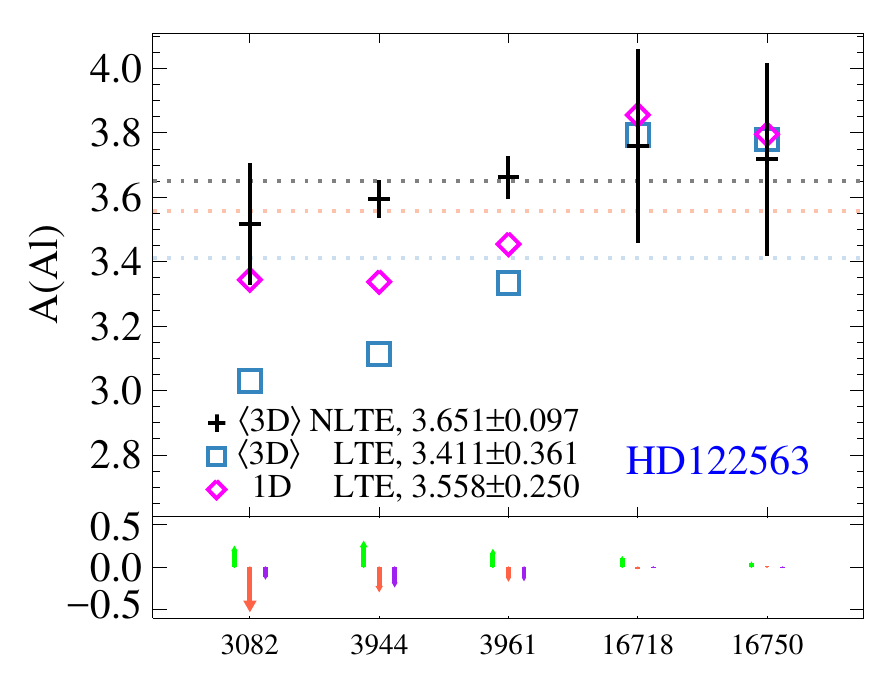}
}
\caption{Abundance analysis of four metal-poor stars, using data from HST/STIS ($\lambda < 3900\,\text{\AA}$), VLT/UVES (optical), and APOGEE/IGRINS (IR).
For \BSSGB, the IR abundances are based on measurements by \citet{afsar_chemical_2016}.
} \label{fig:abundances_mp}
\end{figure*}

Departure coefficients of interpolated \stagger models representing four metal-poor stars are shown in Fig.~\ref{fig:depcoeffs2}, and results of our abundance analyses are shown in Fig.~\ref{fig:abundances_mp}.
At $\FeH = -1$, the spectrum of the metal-poor main-sequence star \BSMS exhibits a reasonably sensitive and unblended resonance line, as well as sufficiently strong optical lines. 
In the spectra of the very metal-poor ($\FeH < -2$) subgiant \BSSGB, the red giant \BSRGB, and the turnoff star \BSTOP, the resonance line is the only unblended optical line of aluminium sufficiently strong for analysis. We also include the resonance line at 3944\,\AA, which is heavily blended with strong lines of CH in carbon-rich stars and therefore not normally recommended for use in abundance analyses.
For diagnostic purposes, we have also analysed HST/STIS UV spectra of all three stars, resulting in as many as a dozen additional lines with little blending for \BSSGB. Additionally, we have analysed near-infrared J-band spectra of \BSRGB. For \BSSGB, we adopt the LTE abundances of infrared lines determined from IGRINS spectra by \citet{afsar_chemical_2016}, but adjust them according to our stellar parameters. 
With the exception of the infrared lines of \BSRGB, all these diagnostic lines exhibit strong deviations from LTE with abundance corrections between $+0.2$ and $+0.7$\,dex, typically with the largest corrections for the 3961\,\ang resonance line.

\subsubsection{\BSMS}
For \BSMS, overionization is the dominant NLTE effect due to the reduced metal opacitites compared to the solar atmosphere, strongly depleting the ground state outside $\log \tau_{500} = 0$.
Resonance scattering causes efficient radiative coupling to the $n \rm s$ and $n \rm d$ states, while infrared transitions among these states are less important. The excited states are essentially thermalized out to $\log \tau_{500} = -1$ due to hydrogen collisional couplings to \ion{Al}{ii} and each other.

Similarly to the Sun, the resulting source function of the 3961\,\ang resonance line is superthermal out to $\log \tau_{500} = -3$, where the wings form, but subthermal at smaller optical depth where the core forms. As the lower state is strongly depleted, the net effect is however a weakening of both core and wings, resulting in a $+0.4$\,dex increase in the inferred abundance from $\Abund{Al} = 5.29$ to $\Abund{Al} = 5.65 \pm 0.08$. 
The average formation depth of all optical lines is less than $\log \tau_{500} = -1$ and the NLTE corrections are subsequently small, at most 0.01\,dex, resulting in a weighted mean abundance of $\Abund{Al} = 5.79\pm0.08$.

\subsubsection{\BSSGB}
For \BSSGB, the further reduction in metallicity and hence UV line opacity as well as lower surface gravity by an order of magnitude results in even stronger overionization of the ground state, compared to \BSMS. 
The effects of photon pumping in the resonance lines are similarly enhanced, where the increasingly superthermal radiation field at shorter wavelengths causes excited states to become increasingly overpopulated. 
The effects are lessened by collisions with hydrogen atoms: charge transfer couples the $3 \rm d$ and $4 \rm p$ states to the \ion{Al}{ii} continuum via charge transfer, making them less depleted. Collisional excitation reduces the overpopulation of $5 \rm p$ and more highly excited states.

The resulting effect on all visible lines in \BSSGB is thus a reduction in line opacity, as well as a superthermal source function, both of which reduce the line strengths.
For the 3961\,\ang resonance line, the abundance correction is $+0.7$\,dex.
As the formation depth moves inward for successively bluer resonance lines, the NLTE correction decreases somewhat but is always larger than $+0.3$\,dex -- on average $+0.38$\,dex. 
While the infrared lines originate in the $\rm 4p$ states which are not as strongly depleted as the ground state, stimulated emission causes a reduction in line opacity as well as enhancement of the superthermal source function, resulting in a large abundance correction of $+0.3$\,dex.
We find good agreement between abundances determined from the UV lines, $\Abund{Al} = 3.48 \pm 0.09$, the 3961\,\ang resonance line, $\Abund{Al} = 3.59\pm0.05$, and the very weak infrared lines, $\Abund{Al} = 3.75 \pm 0.15$.

\subsubsection{\BSTOP}
\BSTOP has similar metallicity to \BSSGB, but its higher $\Teff$ results in stronger overionization of the ground state, with photon pumping in the resonance lines again determining the populations of excited states. 
Efficient electron collisions inside optical depths of $\log \tau_{500} = -1$ lead to near-LTE populations in the excited states, but are inefficient at smaller optical depths where all states become depleted.

NLTE effects on the resonance lines of \BSTOP are thus similar in mechanism and magnitude to \BSSGB: the ground state is depleted, and source functions are superthermal, resulting in positive abundance corrections of $+0.5$\,dex for the 3961\,\ang resonance line and between $+0.2$ and $+0.5$\,dex for the weaker lines in the UV. 
The UV lines indicate $\Abund{Al} = 3.83 \pm 0.13$, in good agreement with the 3961\,\ang resonance line, $\Abund{Al} = 3.94\pm0.02$.

\subsubsection{\BSRGB}
The lower $\Teff$ of \BSRGB results in a significantly smaller UV radiation excess, especially at shorter wavelengths. 
This results in similar effects on the ground state population from overionization and photon pumping, and the resulting overpopulation of highly excited states is smaller than for \BSSGB.
Outside $\log \tau_{500} = -2.5$, collisions have very little influence on all but the most highly excited Rydberg states.

The depletion of the ground state is smaller than for \BSSGB and \BSTOP, and as the source function of the 3961\,\ang resonance line is nearly planckian rather than superthermal at the optical depths where the core forms, the abundance correction is consequently smaller, at $+0.3$\,dex.
The brighter source function of the 3082\,\ang line results in a larger $+0.5$\,dex abundance correction.
The very weak infrared lines form in deep layers where the increased line opacity nearly cancels effects of the darkened source function, such that abundance corrections are small but negative, $-0.05$\,dex. Abundances determined from the UV line, $\Abund{Al} = 3.52 \pm 0.19$, the 3961\,\ang resonance line, $\Abund{Al} = 3.66 \pm 0.07$, and the very weak infrared lines, $\Abund{Al} = 3.74 \pm 0.3$, are in excellent agreement.


\subsection{Grids of NLTE corrections}\label{sect:hrdiag}

\begin{figure*}
	\centerline{\includegraphics[]{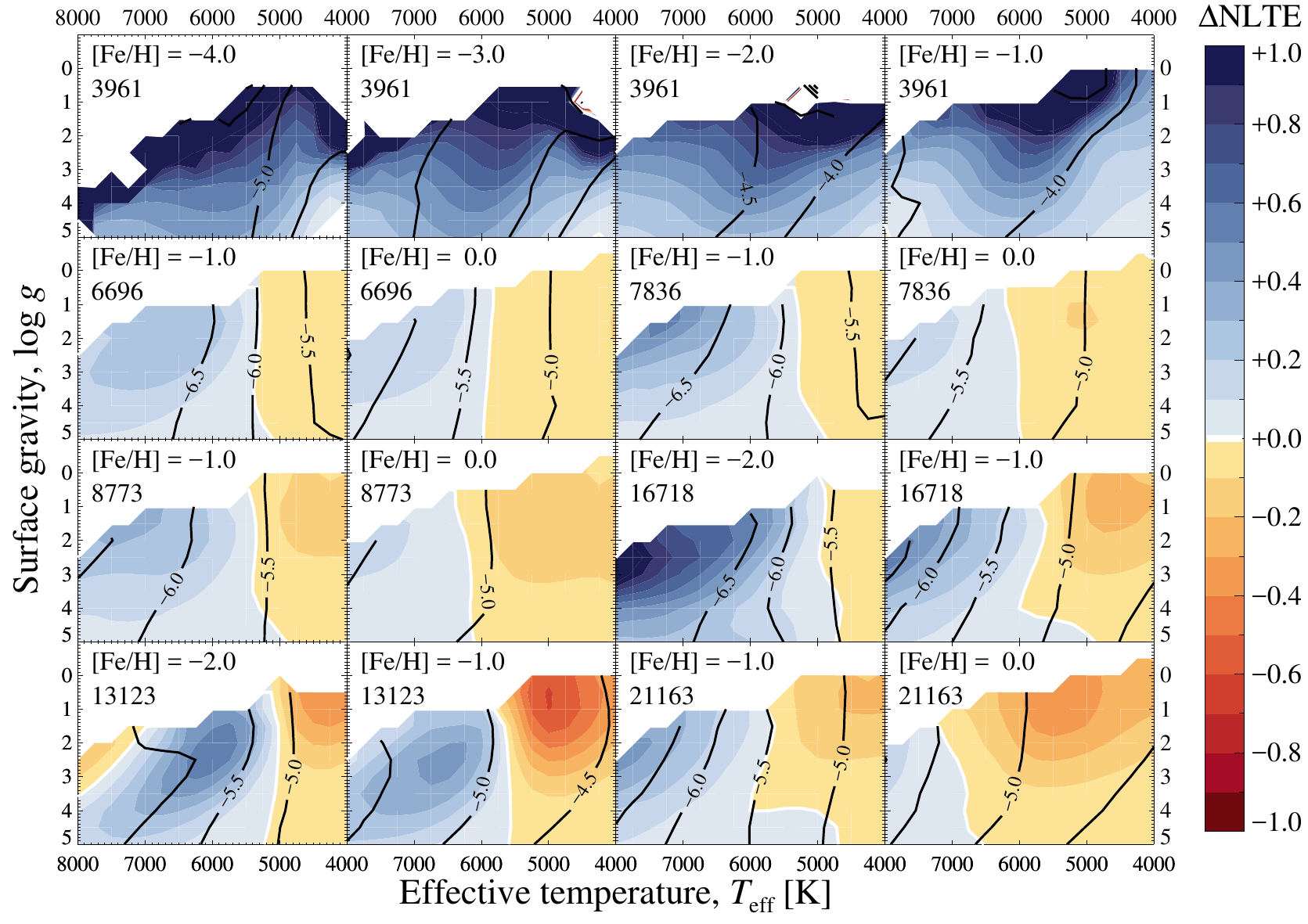}}
	\caption{NLTE abundance corrections for representative lines of aluminium, computed at different metallicities with a NLTE reference abundance of $\eFe{Al} = 0$ for 1D \marcs models. The metallicity and line central wavelength (in units of \ang) are indicated in the top left corner of each panel. Overplotted isocontours indicate the reduced line strength, $\log W_\lambda / \lambda$. The white line delimits positive from negative abundance corrections, and the color scale is given in steps of 0.1\,dex. }
	\label{fig:hrdiag}
\end{figure*}

We have computed NLTE line profiles for atmospheric models in the \avgtd \stagger and 1D \marcs grids with $\Teff > 4000$\,K in a range of abundances between $\eFe{Al} = -2$ and $+2$. 
These calculations assume $\vmic = 1\,\kms$ for dwarfs and $\vmic = 2\,\kms$ for giants, with dwarfs defined as having $\logg \ge 4$ in the 1D calculations but $\logg \ge 3.5$ in \avgtd.
By interpolating LTE equivalent widths onto the NLTE curves of growth, we compute representative NLTE abundance corrections. 

We illustrate these NLTE abundance corrections for models in the 1D \marcs grid in Fig.~\ref{fig:hrdiag} for lines typically used in abundance analyses, at different metallicities, but assuming a NLTE reference abundance of $\eFe{Al} = 0$. The isocontours indicate logarithmic reduced line strengths, $\log W_\lambda / \lambda$, where \eg values of $-6$, $-5$ and $-4$ correspond to line strengths of $W_\lambda = 5$, 50 and 500\,m\ang at a wavelength of $\lambda = 5000$\,\ang. These values represent roughly a typical detection limit at $S/N > 100$, the onset of saturation, and the transition toward the strong line regime, respectively.

Broadly, the 3961\,\ang line always exhibits positive abundance corrections due to depletion of the ground state by overionisation and possibly photon pumping. This causes a reduction in line opacity, often in combination with a superthermal source function. 
The other lines are due to transitions between excited states. The near-IR lines strengthen in cool (or metal-rich) stars compared to the warmer stars, and resonance scattering here has the opposite influence to the UV; photon losses induce an efficient photon suction ladder, resulting in negative abundance corrections. In the warmer (or metal-poor) stars, overionization dominates, resulting in negative abundance corrections.

In practice, the abundance correction depends on the measured abundance, and we stress that the influence of blending lines may affect the abundance correction.
In particular, the line at 3961\,\ang is heavily blended and our synthesis includes background contributions from the \ion{Ca}{ii} H and K lines, as well as the Balmer-$\varepsilon$ line. In metal rich stars, these blending lines are significantly stronger than the aluminium line itself, such that there is no nearby continuum relative to which the spectrum may be normalized for the full equivalent width of the line to be computed. 
As a compromise, we compute the equivalent width of this line over a 4.5\,\ang wide region normalized to the pseudocontinuum computed at line centre, represented by the Al-free flux. This interval is sufficiently short that the background equivalent width nearly cancels -- as the shape of the pseudocontinuum is somewhat parabolic, the Al-free spectrum does have nonzero equivalent width -- but does exclude the contribution of the far wings in very Al-rich spectra.

At extremely low metallicity, the 3961\,\ang resonance line is the only unblended aluminium line visible in the optical region, and its abundance correction is always positive.
At $\eFe{Al} = -4$, the NLTE correction ranges between roughly $+0.1$\,dex on the lower main sequence (MS; \eg $\Teff \approx 4500$\,K, $\logg \approx 4.5$), $+0.5$\,dex near the main-sequence turnoff point (MSTO; \eg $\Teff \approx 6250$\,K, $\logg \approx 4.0$), and $+0.8$\,dex on the RGB (\eg $\Teff \approx 5250$\,K, $\logg \approx 2.0$). 
At $\FeH > -2$, the line may be strongly saturated or even sufficiently strong that the equivalent width is dominated by the wings, which in turn are significantly depressed by the nearby \ion{Ca}{ii} H line, such that the abundance corrections depend on the abundance of Ca and details of the spectrum normalization. 
This influence on the abundance determination is not straightforward, and is in fact different in the NLTE and LTE case such that the NLTE abundance correction depends on the adopted calcium abundance \citep{mashonkina_influence_2016}.
Hence, we urge the use of profile fitting for this line, whenever significant blending is expected. 
Despite these caveats, we note that our grid of equivalent widths indicates an abundance correction of $+0.35$\,dex for \BSMS, where our detailed line profile analyses in NLTE and LTE differed by $+0.36$\,dex.

At higher metallicity, the doublets at 6696, 7836 and 8773\,\ang are commonly used in the literature. We find small negative abundance corrections of at most $-0.04$\,dex on the lower MS. The abundance correction becomes more positive with increasing effective temperature, and ranges between $0.0$ and $+0.1$\,dex on the MSTO, depending on line and metallicity. On the RGB, the lower state $\rm 4s$ or $\rm 3d$ becomes overpopulated, leading to negative abundance corrections between $0.0$ and $-0.15$\,dex. 

The problem with combining results from the optical lines at high metallicity and the resonance lines at low metallicity is clearly illustrated in the NLTE study of \citet[see their Fig.~15]{zhao_systematic_2016}. There, abundance corrections among solar neighbourhood dwarfs vary between $-0.1$ and $+0.1$\,dex for the optical lines at solar metallicity, but reach $+0.5$\,dex when the resonance line is used at low metallicity.

Several strong lines with good atomic data exist in the near-infrared J, H and K bands.
In the H band, the triplet at 16718--16763\,\ang contains a strong line at 16718\,\ang, which is possible to detect down to metallicities of $\FeH = -2.5$ on the RGB, and is covered by \eg the APOGEE survey. At low metallicity, we find that the NLTE correction for this line varies with the line strength, with small but negative abundance corrections of $-0.02$\,dex on the lower MS, but positive corrections of $+0.1$\,dex on the RGB. The correction becomes more positive with increasing $\Teff$, with values of $+0.0$ to $+0.2$\,dex on the MSTO.
At higher metallicity, the abundance correction is unchanged on the lower MS, but decreases to zero on the MSTO. On the RGB, the line becomes strongly saturated, and we find negative abundance corrections between $-0.1$ and $-0.3$\,dex, which vary strongly with the surface gravity. 

For stellar parameters typical of red giant stars in the APOGEE survey, we find that abundance corrections for the 16718\,\ang line vary significantly with stellar parameters. These star-to-star variations amount to roughly 0.2\,dex at any given metallicity. The average correction is $-0.15$\,dex at solar metallicity and $-0.05$\,dex at $\FeH = -2$, such that the prominent downward slope at low metallicity -- seen in \eg, Fig.~14 of \citet{holtzman_abundances_2015} -- flattens somewhat. 
We note however that their illustrated abundance results have been empirically calibrated to remove trends as a function of temperature, which to some extent may already take NLTE effects into account. NLTE corrections should thus preferably be applied to uncalibrated abundances. Additionally, the abundance variations vary rapidly with stellar parameters, and must be applied on a star-by-star basis.

In the J band, the 13150\,\ang line in the 13123--13150\,\ang doublet is possible to detect at metallicities approaching $\FeH = -3.5$. 
As the NLTE effects on this line are delicately balanced by photon suction and the overionization of the ground state, we find that they vary rapidly with stellar parameters.
The abundance correction is more positive at lower metallicity, and varies between $0.00$ and $-0.05$\,dex on the lower MS, between $0.0$ and $+0.2$\,dex on the MSTO, and between $-0.6$ and $-0.1$\,dex on the RGB. 

The K band, which is suitable for use in strongly reddened regions like the inner Galactic bulge, contains a strong line at 21163\,\ang as part of the 21093--21163\,\ang doublet, in a region which is relatively free of telluric absorption. At $\FeH = -1$, we find the abundance correction for this line to vary between $-0.05$\,dex on the lower MS, $+0.05$\,dex on the MSTO, and between $-0.10$\,dex and $-0.25$\,dex on the lower and upper RGB, respectively.

\begin{figure*}
	\centerline{\includegraphics[]{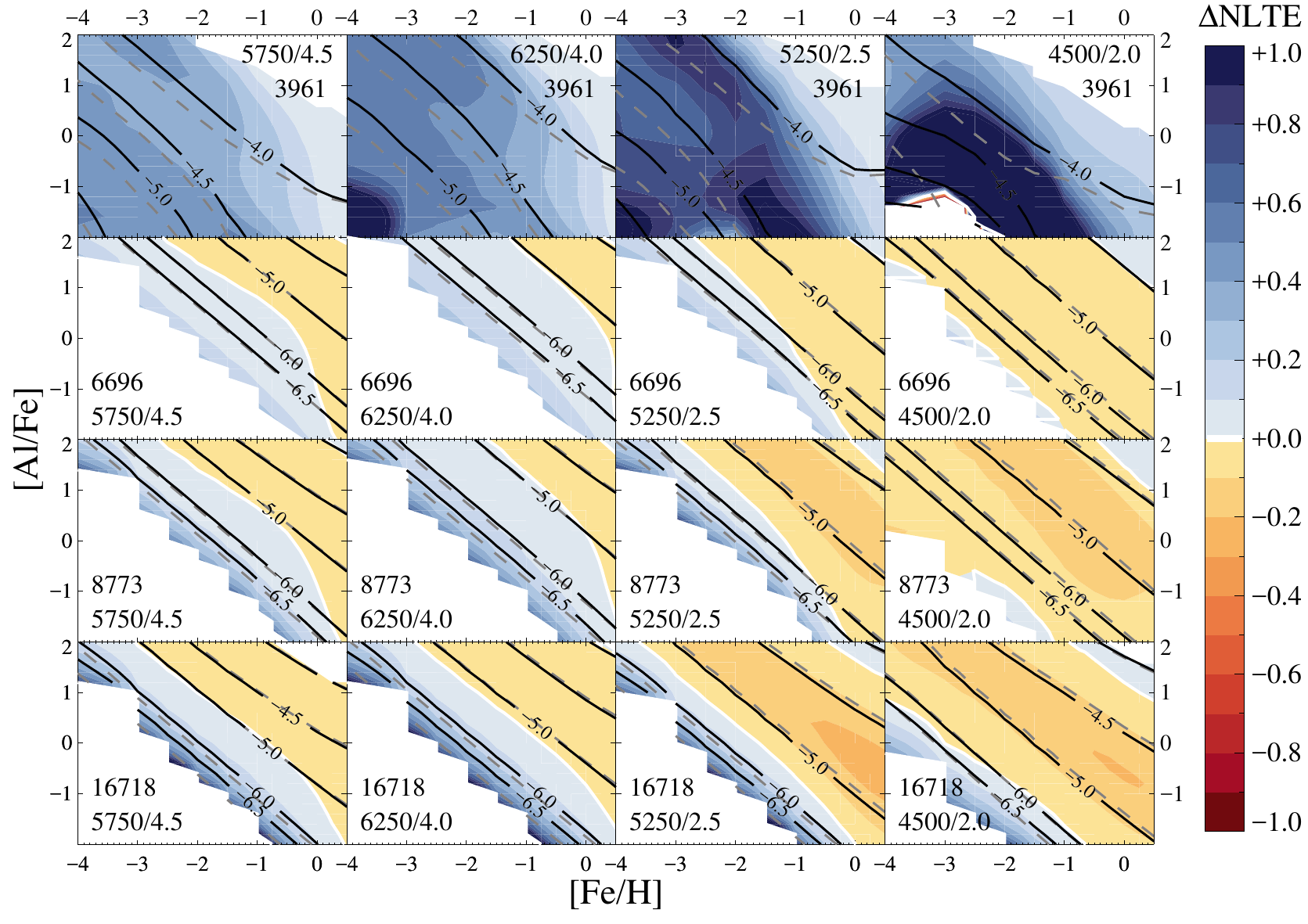}}
	\caption{NLTE abundance corrections for representative lines of aluminium computed in a range of metallicities and LTE aluminium abundances for 1D \marcs models. The stellar parameters $\Teff$ (in units of K) and $\logg$ (in cgs units) and the line central wavelength (in units of \ang) are indicated in each panel. Overplotted isocontours indicate the reduced line strength, $\log W_\lambda / \lambda$, in LTE (dashed lines) and NLTE (labeled solid lines). Each row presents results for a different line, and each column a different stellar model representing, from left to right, solar analogs, the main-sequence turnoff, the lower and the upper RGB.}
	\label{fig:xfe_feh}
\end{figure*}

In addition to significant variation in the NLTE effects as a function of stellar parameters, differential effects may be present even in homogeneous samples of \eg solar analogs which have essentially identical $\Teff$ and $\logg$ but differ in metallicity and abundance. 
The variation of NLTE abundance corrections as a function of metallicity and aluminium abundance is illustrated in Fig.~\ref{fig:xfe_feh}, for four representative lines in four stellar models. While we plot the full range of $\eFe{Al}$ represented in the computed grid, literature abundances are typically confined to the range between $\eFe{Al} = -1$ and $+1$, where low values are found in very metal-poor field stars and high values in second-generation stars in globular clusters \citep[see \eg][]{cayrel_first_2004,carretta_na-o_2009}.

In very metal-poor solar analogs and MSTO models, the NLTE abundance correction for the 3961\,\ang resonance line varies by 0.2\,dex depending on metallicity and aluminium abundance. 
In the very metal-poor models on the lower RGB, the abundance correction may vary both with aluminium abundance at a given metallicity and with metallicity at a given line strength by as much as 0.3\,dex. The large differential effects are due to the varying saturation level of the line, as well as the varying influence of the line itself on the statistical equilibrium.
On the upper RGB, the abundance corrections for spectra with low Al abundances are formally very large (several dex) due to a combination of large NLTE effects weakening the line by a factor of 2 -- due to pumping of the $\rm 4s$ state leading to source function brightening and thus a weakening of the line core -- and strong saturation of the LTE line profile. Computing the abundance correction thus requires extrapolation beyond the available LTE curve of growth. We strongly recommend the use of NLTE line profile matching in these cases.

For all optical and near-infrared lines, the abundance correction becomes more negative with increasing abundance as long as the line is weak. For saturated lines, the negative abundance correction becomes smaller. 
For example, very metal-poor stars ($\FeH = -1$) on the lower RGB may exhibit positive abundance corrections for the 8773\,\ang line of $+0.1$\,dex at low abundance, but negative corrections of $-0.2$\,dex at high abundance, resulting in a compression of the abundance scale, \ie the abundance difference, by as much as 0.3\,dex.

The 6696--6698\,\ang doublet is commonly used in studies of abundance variations in globular clusters. For example, the LTE study by \citet{carretta_na-o_2009} found that Al abundance variations are ubiquitous and typically anticorrelated with Mg. The full extent of abundance variations differ in magnitude from cluster to cluster, scaling broadly with the cluster mass, but we note that the more metal-rich clusters typically exhibit smaller variations than the more metal-poor ones.
The stars in their study vary between $\Teff \approx 4000$\,K and $\logg \approx 1$ at $\FeH = -0.75$, and $\Teff \approx 4500$--5000\,K and $\logg = 1$--2 at $\FeH = -2.5$. 
We typically find NLTE corrections of $-0.10$\,dex for models representing the more metal-rich stars, while for the most metal-poor stars, NLTE effects vary strongly with both stellar parameters and abundance.
For Al-poor stars, these abundance corrections vary between $+0.2$ and $-0.1$\,dex (highest to lowest $\Teff$), while for Al-rich stars the corresponding values are $0.0$ and $-0.1$\,dex.
Hence, while stars in the most metal-rich globular clusters are mainly susceptible to a bias in the inferred abundance of aluminium, the abundance scale among the metal-poor stars is compressed by an amount which varies from cluster to cluster.
As NLTE effects on lines of magnesium and iron also vary with stellar parameters, it is not clear to what extent the analysis of abundance ratios in globular clusters will be affected once NLTE effects are taken into account for all relevant species.

\subsection{Comparison to previous work} \label{sect:previous}

\begin{figure*}
	\centerline{\includegraphics[]{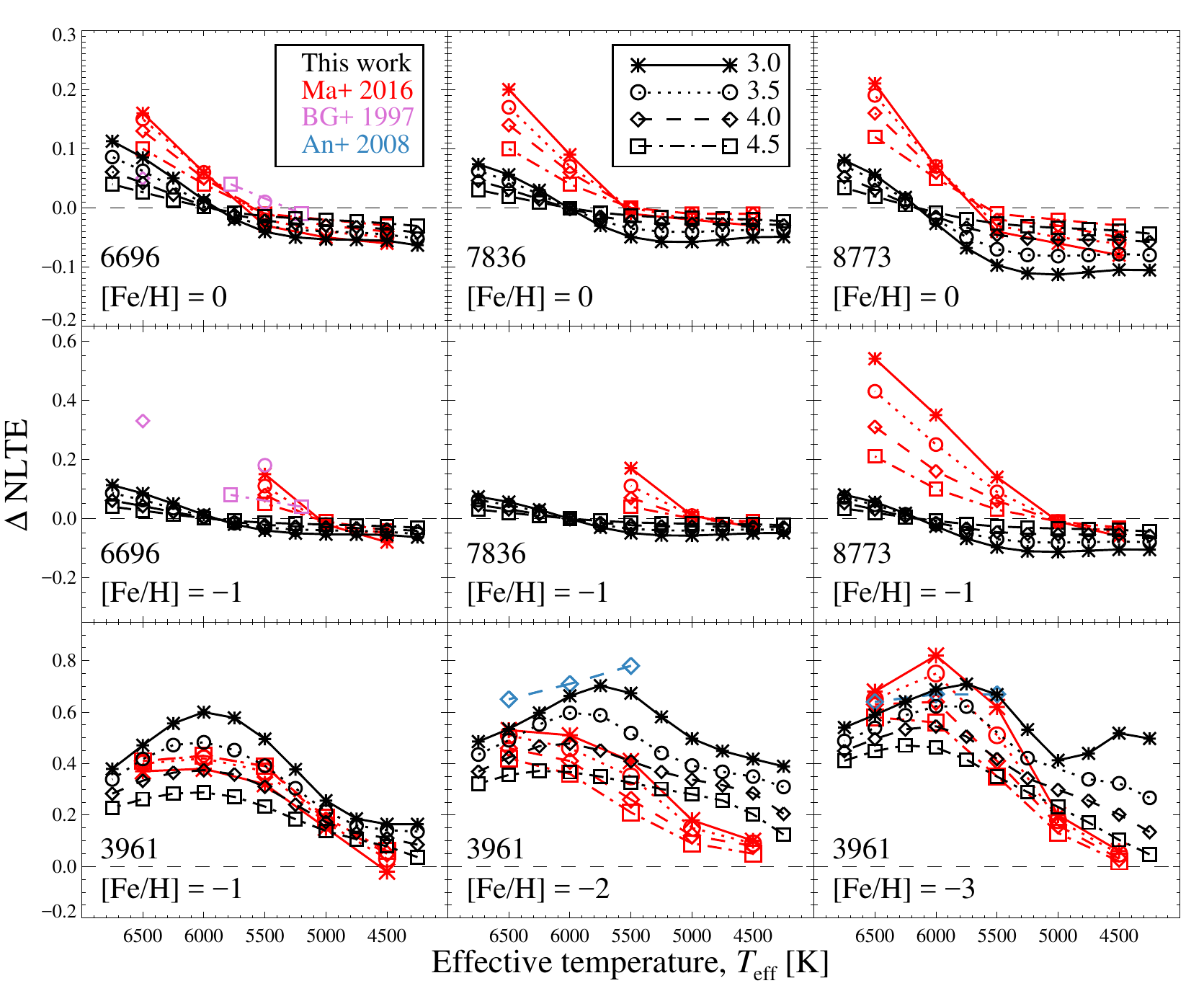}}
	\caption{NLTE abundance corrections for optical lines of aluminium, computed at different metallicities with an LTE reference abundance of $\eFe{Al} = 0$. The metallicity and line central wavelength (in units of \ang) are indicated in each panel. Results from this work (shown in black) are compared to results from \citet[shown in red]{mashonkina_influence_2016}, \citet[shown in purple]{baumuller_aluminium_1997}, and \citet[shown in blue]{andrievsky_nlte_2008}. 
}
	\label{fig:cf_lit}
\end{figure*}

In Fig.~\ref{fig:cf_lit}, we compare our abundance corrections based on equivalent widths (EW) at a fixed LTE abundance $\eFe{Al} = 0$, to those presented by \citet{mashonkina_influence_2016}, who use the same hydrogen collisional data for low-lying states \citep[\ie][]{belyaev_inelastic_2013}. Notably, our abundance corrections for the optical lines are more negative by up to 0.1\,dex. However, the difference in abundance corrections between the hot and cool end of the temperature scale are rather similar, such that the difference mainly affects the resulting absolute abundance scale rather than abundance differences. Additionally, the largest differences are found for rather extreme stellar parameters, \eg $\Teff > 6000$\,K, $\logg = 3$, which would represent \eg blue stragglers or pre-main sequence stars rather than typical field stars. For stellar parameters representative of typical field stars, differences are smaller.
The corrections from \citet{baumuller_aluminium_1997} also differ by $0.0$ to $+0.1$\,dex from ours, but are largely consistent with those of \citet{mashonkina_influence_2016}. 

For the resonance line, EW-based abundance corrections are not straightforward to compute or compare. 
Due to the significant influence of blends and the very large NLTE effects affecting both line strengths and shapes, we recommend that these abundance corrections are used only as a guide.
Differences in the abundance corrections may be due to systematics unrelated to the NLTE modeling, such as differences in background opacities, line broadening, and the method of normalization or selected wavelength interval over which EWs are computed. The effects are most significant when the lines are saturated (sensitive to, \eg, line broadening), very weak (blending and normalization) or very strong (EW wavelength interval).
With these shortcomings in mind, we note that our modeling indicates a stronger variation of NLTE corrections with $\logg$ at $\FeH = -1$, and significantly larger corrections for $\Teff < 5000$\,K, as compared to \citet{mashonkina_influence_2016}.
The most significant differences are found for the resonance line at low $\logg$, $\Teff$ and $\FeH$, \eg, $\Teff = 4500$\,K, $\logg = 3$, $\FeH \le -2$, where our abundance corrections are very large, $+0.6$\,dex, while they find negligibly small corrections of the order 0.05\,dex. As the line is strongly saturated for these parameters, it is possible that the difference is due to systematics unrelated to the NLTE effect.
We have shown for \BSRGB, which has similar $\Teff$ but lower $\logg$, that our large predicted abundance correction of $+0.4$\,dex matches well the abundances determined in the UV and IR. 
For the resonance lines, we also compare to the abundance corrections of \citet{andrievsky_nlte_2008} for dwarf stars. They predict more positive abundance corrections than ours by 0.1--0.3\,dex. As their modeling technique is similar to ours, this mainly indicates the influence of the new collisional rates.

In this work, we have tested our NLTE modeling on a set of benchmark stars, and find that different lines agree to within measurement uncertainties. 
Comparing our theoretical abundance corrections for stellar parameters representative of typical field stars, differences to other studies may be as large as $-0.1$\,dex for the optical lines and $+0.3$\,dex for the resonance lines, and are likely due mainly to the use of different sources for electron collisional rates.
For the optical lines, differences are similar for any given set of stellar parameters, and would thus mainly result in a shift of the abundance scale.
For the resonance line, our corrections imply higher abundances for cool dwarfs and giants, but largely consistent results for warmer dwarfs (to within 0.1\,dex) with no clear systematic offset. Differences for the cool dwarfs may however be due to systematics related to how the abundance correction is estimated, rather than modeling differences.



\section{Conclusions} \label{sect:conclusions}

We have presented a new aluminium model atom for use in NLTE abundance analyses. 
The model uses accurate collisional rates for interactions with both hydrogen atoms and electrons, with no need for empirically calibrated scaling factors.
We also utilize new parameters for collisional broadening by hydrogen atoms, commonly known as van der Waals broadening. These were computed for this work, so that these are now available for every line commonly used in the literature.
We have performed extensive calculations using grids of model atmospheres, utilizing both \avgtd \stagger and 1D hydrostatic \marcs models. For the Sun, we have also for the first time performed full 3D NLTE calculations. 

We have verified the accuracy of our NLTE approach using solar flux and intensity spectra, as well as high-resolution spectra of standard stars.
We find that most line profiles are accurately reproduced, and that the line-to-line scatter is improved when taking NLTE effects into account.
For the Sun, 3D NLTE synthesis reproduces the observed center-to-limb variations in the 7835\,\ang line for disk positions between $\mu = 0.15$ and $\mu=1$, with deviations less than 0.03\,dex. 
Abundance differences between 3D NLTE and \avgtd NLTE synthesis are less than 0.01\,dex at disk centre, but amount to 0.15\,dex at $\mu = 0.15$. 

For the Sun, Arcturus and Pollux, we reproduce the mid-IR emission line at 12.3\,\micron rather well (Figs.~\ref{fig:MIRsun} and \ref{fig:MIRlines}). While observations of this weak line are challenging due to \eg thermal noise and the presence of telluric lines, we note that we appear to slightly but systematically underestimate the line strength. 
As found in previous work \citep{baumuller_line_1996}, the strong line at 1.31\,\micron (Fig.~\ref{fig:NIRlines}) is deeper than predicted in the solar spectrum.
If this shortcoming is due to errors in the NLTE calculations, it would imply that bulk transition rates due to collisions with hydrogen may be in error by at most a factor of ten.
In Arcturus, the line is instead slightly more shallow than predicted, which could likely be a hydrodynamical effect rather than a shortcoming of the NLTE modeling.

Our 3D NLTE analysis of solar disk-centre spectra yields an abundance of $\Abund{Al} = 6.43 \pm 0.03$ (including systematic errors), which is in exact agreement with the meteoritic abundance, $\Abund{Al} = 6.43 \pm 0.01$ (\citealt{lodders_abundances_2009}, renormalized to the abundance of Si derived by \citealt{scott_elemental_2015-1} and \citealt{amarsi_solar_2017}).

In our analysis of Arcturus, we find abundance corrections of $-0.1$ to $-0.2$\,dex for optical lines, and as large as $-0.3$\,dex for the saturated near-infrared lines, resulting in very good agreement.
We also analysed four very metal-poor stars, and find large abundance corrections for the 3961\,\ang resonance line ranging between $+0.3$ and $+0.7$\,dex.
The resulting abundances agree well with lines in the optical and near-infrared, as well as higher-order resonance lines in the UV.

Our grids of abundance corrections indicate that NLTE effects may vary rapidly with varying stellar parameters and must be applied on a star-by-star basis.
For example, abundance corrections for the near-infrared lines at 1.67\,\micron applicable to stars in the APOGEE survey vary from star to star by 0.2\,dex at any given metallicity. The abundance corrections are on average $-0.15$\,dex at solar metallicity but weaken to $-0.05$\,dex at low metallicity. 
This means that NLTE effects affect not only the average trend of aluminium abundances with metallicity, but also the inferred dispersion, with implications for the formation of the Galactic halo and the possible bimodality of the Galactic thin and thick disks.
Similar effects are expected for studies of internal pollution of  globular clusters, where we find that abundances in metal-rich red giants are likely overestimated by 0.1\,dex. Additionally, we find that the full extent of aluminium abundance variations among metal-poor stars is likely overestimated by a similar amount.

Given these very encouraging results, we are confident that our abundance corrections are sufficiently reliable for scientific work.
We therefore make available extensive grids of abundance corrections for lines suitable for abundance analyses in the optical and near-infrared via the INSPECT database\footnote{Available online at \url{http://inspect-stars.com/}}. While we illustrate representative corrections for the 3961\,\ang line, we do not tabulate these corrections as they may be misleading due to the significant influence of blending with \ion{Ca}{ii}. Instead, we urge the use of profile fitting, either via precomputed line profiles or by direct synthesis using grids of departure coefficients -- both of which are available on request.

\begin{acknowledgements}
The authors wish to thank Anish Amarsi for making his version of the \multitd code available to us along with tailored background line opacities, 
Paul Barklem for providing broadening parameters computed for this work,
Thomas Ayres for processing UV HST/STIS spectra of \BSSGB and Ruth Peterson for bringing them to our attention, 
Ken Hinkle for providing the near-IR spectra of Pollux, and 
Jon Sundqvist for providing the mid-IR 12.3\,\micron spectra. 
TN acknowledges support from the Swedish National Space Board (Rymdstyrelsen), 
and funding from Australian Research Council (grant DP150100250).
KL acknowledges funds from the Alexander von Humboldt Foundation in the framework of the Sofja Kovalevskaja Award endowed by the Federal Ministry of Education and Research as well as funds from the Swedish Research Council (Grant nr. 2015-00415\_3) and Marie Sklodowska Curie Actions (Cofund Project INCA 600398).
The computations were performed on resources provided by the Swedish National Infrastructure for Computing (SNIC) at High Performance Computing Center North (HPC2N) under projects SNIC2015/1-309 and SNIC2016/1-400.
\end{acknowledgements}

\bibliography{nordlander_etal_al_nlte,everything}

\appendix

\section{Line data}
\begin{table*}
\caption{Line data for spectral lines used in the abundance analyses} 
\label{tbl:linedata}
\centering 
\begin{tabular}{r r@{}l r@{}l c r@{}l c rl}
\hline\hline \noalign{\smallskip}
Wavelength & \multicolumn2c{Transition} & \multicolumn2c{$J_\text{low}$--$J_\text{upp}$} & $E_\text{low}$ & \multicolumn 2 c {$\log gf$ \tablefootmark a} & \multicolumn 1 c {Ref}  & \multicolumn 2 c {$\Gamma_6$} \\ 
\multicolumn 1 c {$[\text{\AA}]$} & &&  && [eV] & \multicolumn 2 c {dex} &  & $\sigma$ $[\text{a.u.}]$ & \multicolumn 1 c {$\alpha$} \\
\hline \noalign{\smallskip}
     2103.017&      3p\,$^2$P$^\text o$&\,--\,14d\,$^2$D               &3/2&--5/2 &  0.014&$  -1.619$&              &         1 &         & \\ 
     2103.024&      3p\,$^2$P$^\text o$&\,--\,14d\,$^2$D               &3/2&--3/2 &  0.014&$  -2.566$&              &         1 &         & \\ 
     2118.312&      3p\,$^2$P$^\text o$&\,--\,11d\,$^2$D               &1/2&--3/2 &  0.000&$  -1.557$& \,$\pm\,0.10$&         2 &         & \\ 
     2134.732&      3p\,$^2$P$^\text o$&\,--\,10d\,$^2$D               &3/2&--5/2 &  0.014&$  -1.129$& \,$\pm\,0.10$&         2 &         & \\ 
     2134.760&      3p\,$^2$P$^\text o$&\,--\,10d\,$^2$D               &3/2&--3/2 &  0.014&$  -2.090$& \,$\pm\,0.18$&         2 &         & \\ 
     2145.555&      3p\,$^2$P$^\text o$&\,--\,9d\,$^2$D                &1/2&--3/2 &  0.000&$  -1.244$& \,$\pm\,0.10$&         2 &         & \\ 
     2174.027&      3p\,$^2$P$^\text o$&\,--\,8d\,$^2$D                &3/2&--5/2 &  0.014&$  -0.824$& \,$\pm\,0.10$&         2 &         & \\ 
     2174.113&      3p\,$^2$P$^\text o$&\,--\,8d\,$^2$D                &3/2&--3/2 &  0.014&$  -1.778$& \,$\pm\,0.30$&         2 &         & \\ 
     2204.590&      3p\,$^2$P$^\text o$&\,--\,8s\,$^2$S                &3/2&--1/2 &  0.014&$  -2.293$& \,$\pm\,0.10$&         2 &         & \\ 
     2204.660&      3p\,$^2$P$^\text o$&\,--\,7d\,$^2$D                &1/2&--3/2 &  0.000&$  -0.895$& \,$\pm\,0.06$&         2 & 2340&1.07 \\ 
     2210.046&      3p\,$^2$P$^\text o$&\,--\,7d\,$^2$D                &3/2&--5/2 &  0.014&$  -0.641$& \,$\pm\,0.06$&         2 & 2340&1.07 \\ 
     2210.130&      3p\,$^2$P$^\text o$&\,--\,7d\,$^2$D                &3/2&--3/2 &  0.014&$  -1.595$& \,$\pm\,0.06$&         2 & 2340&1.07 \\ 
     2269.096&      3p\,$^2$P$^\text o$&\,--\,6d\,$^2$D                &3/2&--5/2 &  0.014&$  -0.454$& \,$\pm\,0.04$&       2,3 &2084&0.960 \\ 
     2269.220&      3p\,$^2$P$^\text o$&\,--\,6d\,$^2$D                &3/2&--3/2 &  0.014&$  -1.409$& \,$\pm\,0.06$&         3 &2084&0.960 \\ 
     2367.052&      3p\,$^2$P$^\text o$&\,--\,5d\,$^2$D                &1/2&--3/2 &  0.000&$  -0.592$& \,$\pm\,0.06$&         2 &         & \\ 
     2372.070&      3p\,$^2$P$^\text o$&\,--\,6s\,$^2$S                &1/2&--1/2 &  0.000&$  -2.012$& \,$\pm\,0.10$&         2 &         & \\ 
     2373.124&      3p\,$^2$P$^\text o$&\,--\,5d\,$^2$D                &3/2&--5/2 &  0.014&$  -0.337$& \,$\pm\,0.04$&         2 &         & \\ 
     2373.349&      3p\,$^2$P$^\text o$&\,--\,5d\,$^2$D                &3/2&--3/2 &  0.014&$  -1.291$& \,$\pm\,0.06$&         2 &         & \\ 
     2567.984&      3p\,$^2$P$^\text o$&\,--\,4d\,$^2$D                &1/2&--3/2 &  0.000&$  -1.119$& \,$\pm\,0.04$&         3 &1385&0.249 \\ 
     2575.094&      3p\,$^2$P$^\text o$&\,--\,4d\,$^2$D                &3/2&--5/2 &  0.014&$  -0.668$& \,$\pm\,0.06$&       2,3 &1387&0.249 \\ 
     2575.393&      3p\,$^2$P$^\text o$&\,--\,4d\,$^2$D                &3/2&--3/2 &  0.014&$  -1.623$& \,$\pm\,0.06$&         3 &1385&0.249 \\ 
     2652.484&      3p\,$^2$P$^\text o$&\,--\,5s\,$^2$S                &1/2&--1/2 &  0.000&$  -1.523$& \,$\pm\,0.04$&         4 &2150&0.307 \\ 
     2660.392&      3p\,$^2$P$^\text o$&\,--\,5s\,$^2$S                &3/2&--1/2 &  0.014&$  -1.219$& \,$\pm\,0.04$&         4 &2150&0.307 \\ 
     3082.153&      3p\,$^2$P$^\text o$&\,--\,3d\,$^2$D                &1/2&--3/2 &  0.000&$  -0.476$&\,$\pm\,0.025$&       2,3 & 576&0.292 \\ 
     3092.710&      3p\,$^2$P$^\text o$&\,--\,3d\,$^2$D                &3/2&--5/2 &  0.014&$  -0.202$&\,$\pm\,0.025$&       2,5 & 576&0.292 \\ 
     3092.839&      3p\,$^2$P$^\text o$&\,--\,3d\,$^2$D                &3/2&--3/2 &  0.014&$  -1.175$& \,$\pm\,0.04$&         2 & 576&0.292 \\ 
     3944.006&      3p\,$^2$P$^\text o$&\,--\,4s\,$^2$S                &1/2&--1/2 &  0.000&$  -0.635$&\,$\pm\,0.025$&       2,5 & 655&0.243 \\ 
     3961.520&      3p\,$^2$P$^\text o$&\,--\,4s\,$^2$S                &3/2&--1/2 &  0.014&$  -0.333$&\,$\pm\,0.025$&         2 & 655&0.243 \\ 
     6696.015&                4s\,$^2$S&\,--\,5p\,$^2$P$^\text o$      &1/2&--3/2 &  3.143&$  -1.569$& \,$\pm\,0.06$&         2 &1860&0.226 \\ 
     6698.672&                4s\,$^2$S&\,--\,5p\,$^2$P$^\text o$      &1/2&--1/2 &  3.143&$  -1.870$& \,$\pm\,0.06$&         2 &1860&0.226 \\ 
     7835.309&                3d\,$^2$D&\,--\,6f\,$^2$F$^\text o$      &3/2&--5/2 &  4.022&$  -0.689$& \,$\pm\,0.04$&         2 & 3850&1.60 \\ 
     7836.134&                3d\,$^2$D&\,--\,6f\,$^2$F$^\text o$      &5/2&--5/2 &  4.022&$  -1.834$& \,$\pm\,0.06$&         2 & 3850&1.60 \\ 
     7836.134&                3d\,$^2$D&\,--\,6f\,$^2$F$^\text o$      &5/2&--7/2 &  4.022&$  -0.534$&\,$\pm\,0.025$&         2 & 3850&1.60 \\ 
     8772.866&                3d\,$^2$D&\,--\,5f\,$^2$F$^\text o$      &3/2&--5/2 &  4.022&$  -0.349$&\,$\pm\,0.025$&         2 &2983&0.344 \\ 
     8773.895&                3d\,$^2$D&\,--\,5f\,$^2$F$^\text o$      &5/2&--5/2 &  4.022&$  -1.495$& \,$\pm\,0.06$&         2 &2983&0.344 \\ 
     8773.895&                3d\,$^2$D&\,--\,5f\,$^2$F$^\text o$      &5/2&--7/2 &  4.022&$  -0.192$&\,$\pm\,0.025$&         2 &2983&0.344 \\ 
     8912.900&      4p\,$^2$P$^\text o$&\,--\,6d\,$^2$D                &1/2&--3/2 &  4.085&$  -1.963$& \,$\pm\,0.06$&         2 &         & \\ 
     8923.555&      4p\,$^2$P$^\text o$&\,--\,6d\,$^2$D                &3/2&--5/2 &  4.087&$  -1.709$& \,$\pm\,0.06$&         2 &         & \\ 
     8925.503&      4p\,$^2$P$^\text o$&\,--\,6d\,$^2$D                &3/2&--3/2 &  4.087&$  -2.663$&              &          2&         & \\ 
    10768.363&      4p\,$^2$P$^\text o$&\,--\,5d\,$^2$D                &1/2&--3/2 &  4.085&$  -2.020$& \,$\pm\,0.06$&         2 &         & \\ 
    10782.046&      4p\,$^2$P$^\text o$&\,--\,5d\,$^2$D                &3/2&--5/2 &  4.087&$  -1.764$& \,$\pm\,0.06$&         2 &         & \\ 
    10786.770&      4p\,$^2$P$^\text o$&\,--\,5d\,$^2$D                &3/2&--3/2 &  4.087&$  -2.719$& \,$\pm\,0.10$&         2 &         & \\ 
    10872.975&      4p\,$^2$P$^\text o$&\,--\,6s\,$^2$S                &1/2&--1/2 &  4.085&$  -1.326$& \,$\pm\,0.06$&         2 &         & \\ 
    10891.732&      4p\,$^2$P$^\text o$&\,--\,6s\,$^2$S                &3/2&--1/2 &  4.087&$  -1.027$& \,$\pm\,0.04$&         2 &         & \\ 
    13123.416&                4s\,$^2$S&\,--\,4p\,$^2$P$^\text o$      &1/2&--3/2 &  3.143&$   0.219$&\,$\pm\,0.013$&         2 & 815&0.223 \\ 
    13150.753&                4s\,$^2$S&\,--\,4p\,$^2$P$^\text o$      &1/2&--1/2 &  3.143&$  -0.083$&\,$\pm\,0.013$&         2 & 815&0.223 \\ 
    16718.974&      4p\,$^2$P$^\text o$&\,--\,4d\,$^2$D                &1/2&--3/2 &  4.085&$   0.220$&\,$\pm\,0.013$&         2 &1147&0.311 \\ 
    16750.519&      4p\,$^2$P$^\text o$&\,--\,4d\,$^2$D                &3/2&--5/2 &  4.087&$   0.474$&\,$\pm\,0.013$&         2 &1147&0.311 \\ 
    16763.369&      4p\,$^2$P$^\text o$&\,--\,4d\,$^2$D                &3/2&--3/2 &  4.087&$  -0.480$&\,$\pm\,0.025$&         2 &1147&0.311 \\ 
    21093.082&      4p\,$^2$P$^\text o$&\,--\,5s\,$^2$S                &1/2&--1/2 &  4.085&$  -0.398$&\,$\pm\,0.025$&         4 &1838&0.279 \\ 
    21163.803&      4p\,$^2$P$^\text o$&\,--\,5s\,$^2$S                &3/2&--1/2 &  4.087&$  -0.093$&\,$\pm\,0.025$&         4 &1838&0.279 \\ 
    21208.171&      4f\,$^2$F$^\text o$&\,--\,7g\,$^2$G                &7/2&--9/2 &  5.123&$  -0.307$&              &          2& 3260&1.68 \\ 
    21208.171&      4f\,$^2$F$^\text o$&\,--\,7g\,$^2$G                &5/2&--7/2 &  5.123&$  -0.461$&              &          2& 3260&1.68 \\ 
    21208.191&      4f\,$^2$F$^\text o$&\,--\,7g\,$^2$G                &7/2&--7/2 &  5.123&$  -1.608$&              &          2& 3260&1.68 \\ 
   123349.6\tablefootmark b %
             &      6h\,$^2$H$^\text o$&\,--\,7i\,$^2$I                && &  5.614&$   1.647$&              &          6& 4830&1.84 \\ 
\hline
\end{tabular}
\tablefoot{
Only lines used in the spectrum analyses are listed. 
\tablefoottext{a}{Uncertainties (90\,\%) to the transition probabilities are taken from \citet{kelleher_atomic_2008}, when available.}
\tablefoottext{b}{Multiplet components with identical wavelength have been merged. We adopt the wavelength measured from the solar spectrum by \citet{brault_solar_1983}, and the oscillator strength has been adjusted accordingly.}
}
\tablebib{
(1) \citet[][]{wiese_atomic_1969};
(2)~TOPbase: \citet{mendoza_atomic_1995} and C. Mendoza, W. Eissner, M. Le Dourneuf, and C. J. Zeippen (unpublished);
(3) \citet{davidson_lifetimes_1990};
(4) \citet{vujnovic_absolute_2002};
(5) \citet{tachiev_almchf_2002};
(6) Kurucz (2012, online data; \url{http://kurucz.harvard.edu/atoms/1300/}).
}
\end{table*}

\begin{table}
\caption{Hyperfine constants for \ion{Al}i.}
\label{tbl:hfs}
\centering
\begin{tabular}{l c cc c}
\hline\hline \noalign{\smallskip}
State & $n$ & \multicolumn1c{$A$} & \multicolumn1c{$B$} & Reference \\
& & \multicolumn1c{[MHz]} & \multicolumn1c{[MHz]} \\
\hline \noalign{\smallskip}
$^2 S_{1/2}$  &  4 &  431.84  &   0    & 1 \\ 
              &  5 & (145.5)  &   0    & $A \propto n_\text{eff}^{-3}$ \\ 
              &  6 &  (59.8)  &   0    \\
              &  7 &  (31.5)  &   0    \\
              &  8 &  (18.6)  &   0    \\
              &  9 &  (11.9)  &   0    \\
              & 10 &  (8.08)  &   0    \\
$^2 P_{1/2}$  &  3 & 498.33   &   0    & 2 \\ 
              &  4 &  58.28   &   0    & 2 \\ 
              &  5 &  20      &   0    & 3 \\ 
              &  6 &   (8.03) &   0    & $A \propto n_\text{eff}^{-3.6}$ \\
              &  7 &   (4.01) &   0    & \\
              &  8 &   (2.24) &   0    & \\
$^2 P_{3/2}$  &  3 &  93.76   &  19.12 & 1 \\ 
              &  4 &  23.12   &   0    & 2 \\ 
              &  5 &   (10.0) &   0    & $A \propto n_\text{eff}^{-2.5}$ \\ %
              &  6 &   5.7    &   0.5  & 4 \\ 
              &  7 &   3.3    &   0.3  & 4 \\ 
              &  8 &   2.1    &   0.2  & 4 \\ 
$^2 D_{3/2}$  &  3 & $-99$    & $-13$  & 5,6 \\ 
              &  4 & $-72$    &   0    & 5,6 \\ 
              &  5 & ($-54$)  &   0    & $A \propto n_\text{eff}^{-1.25}$ \\ 
$^2 D_{5/2}$  &  3 & 182      &  22    & 6 \\ 
              &  4 & 204      &   0    & 6 \\ 
              &  5 & 162      &   0    & 5 \\ 
\hline
\end{tabular}
\tablefoot{
Values in parentheses are based on the indicated extrapolations, fitted to the experimental data.
}
\tablebib{
(1) \citet{nakai_hyperfine_2007};
(2) \citet{sur_comparative_2005};
(3) \citet{belfrage_hyperfine_1984};
(4) \citet{jonsson_hyperfine_1984};
(5) \citet{chang_energy_1990};
(6) \citet{falkenburg_perturbed_1979}.
}
\end{table}

Line data employed in the NLTE calculations as well as our abundance analyses are given in Table~\ref{tbl:linedata}.
The VdW $\Gamma_6$ cross-section $\alpha$ and velocity parameter $\sigma$ are often tabulated as $\text{int}(\sigma) + \alpha$. We note that this notation is not possible when $\alpha > 1$, which often occurs for transitions where the upper state is highly excited. In these cases, we recommend to use $\alpha = 0.999$ when photospheric temperatures are similar to the reference temperature 5000\,K.
The $A$ and $B$ constants for hyperfine structure used in this work are listed in Table~\ref{tbl:hfs}. Since $^{27}$Al is the only stable isotope of aluminium, its nuclear spin $I = 5/2$ was used in the calculations. We evaluate the strengths of hyperfine line components following the formulae given by \citet{prochaska_galactic_2000}.

\end{document}